\documentclass[11pt]{article}
\usepackage[utf8]{inputenc}
\usepackage{caption}
\usepackage{subcaption}
\usepackage{braket,slashed,bm}
\usepackage{jheppub}
\usepackage{simplewick}
\usepackage{array,multirow}
\usepackage{amsmath} 
\usepackage[normalem]{ulem}
\usepackage{calligra}
\usepackage{floatrow}
\DeclareUnicodeCharacter{2212}{\ensuremath{-}}
\DeclareMathAlphabet{\mathpzc}{OT1}{pzc}{m}{it}
\usepackage{mathrsfs}
\usepackage{tikz}
\usepackage{rotating}
\usepackage{pgf}

\usepackage{subcaption}
\usepackage{lscape}
\usepackage{graphicx}
\usepackage{booktabs}
\newfloatcommand{capbtabbox}{table}[][\FBwidth]

\usepackage{amsmath}
\usepackage{graphicx,bm}

\allowdisplaybreaks
\begin{document}

\title{
Exploring Mirror Twin Higgs Cosmology with Present and Future Weak Lensing Surveys
}
\author[a,b]{Lei Zu\footnote{\label{contribute}Contributed Equally},}
\author[a,b]{Chi Zhang\footnote{\label{contribute}Contributed Equally},}
\author[a,b]{Hou-Zun Chen,}
\author[a,b]{Wei Wang,}
\author[a,b]{Yue-Lin Sming Tsai\footnote{Corresponding Author: smingtsai@pmo.ac.cn},}
\author[c]{Yuhsin Tsai\footnote{Corresponding Author: ytsai3@nd.edu},}
\author[d]{Wentao Luo,}
\author[a,b]{Yi-Zhong Fan}

\affiliation[a]{Key Laboratory of Dark Matter and Space Astronomy, 
Purple Mountain Observatory, Chinese Academy of Sciences, Nanjing 210033, China}
\affiliation[b]{School of Astronomy and Space Science, University of Science and Technology of China, Hefei, Anhui 230026, China}
\affiliation[c]{Department of Physics, University of Notre Dame, IN 46556, USA}
\affiliation[d]{Department of Astronomy, School of Physical Sciences, University of Science and Technology of China, 
Hefei, Anhui 230026,China}

\abstract{
We explore the potential of precision cosmological data to study non-minimal dark sectors by updating the cosmological constraint on the mirror twin Higgs model (MTH). The MTH model addresses the Higgs little hierarchy problem by introducing dark sector particles. 
In this work, we perform a Bayesian global analysis that includes the latest cosmic shear measurement from the DES three-year survey and the Planck CMB and BAO data. 
In the early Universe, the mirror baryon and mirror radiation behave as dark matter and dark radiation, 
and their presence modifies the Universe's expansion history. 
Additionally, the scattering between mirror baryon and photon generates the dark acoustic oscillation process, suppressing the matter power spectrum from the cosmic shear measurement. 
We demonstrate how current data constrain these corrections to the $\Lambda$CDM cosmology and find that for a viable solution to the little hierarchy problem, the proportion of MTH dark matter cannot exceed about $30\%$ of the total dark matter density, unless the temperature of twin photon is less than $30\%$ of that of the standard model photon. While the MTH model is presently not a superior solution to the observed $H_0$ tension compared to the $\Lambda$CDM+$\Delta N_{\rm eff}$ model, we demonstrate that it has the potential to alleviate both the $H_0$ and $S_8$ tensions, especially if the $S_8$ tension persists in the future and approaches the result reported by the Planck~SZ~(2013) analysis. In this case,  the MTH model can relax the tensions while satisfying the DES power spectrum constraint up to $k \lesssim 10~h\rm {Mpc}^{-1}$. If the MTH model is indeed accountable for the $S_8$ and $H_0$ tensions, we show that the future China Space Station Telescope (CSST) can determine the twin baryon abundance with a $10\%$ level precision. 
}

\date{\today}

\maketitle

\section{Introduction} 
The mirror twin Higgs (MTH) model~\cite{Chacko_2006,Chacko2_2006} is a well-motivated scenario that solves the Higgs little hierarchy problem without being subject to severe constraints from measurements at the Large Hadron Collider (LHC). 
The model features a softly broken mirror symmetry, and each Standard Model (SM) particle has a mirror partner carrying the same gauge and Yukawa couplings to the SM but only interacts with twin particles at low energy. 
The presence of mirror baryons, photons, and neutrinos predicted by the model leads to new cosmological signatures with length scales ranging from galaxy formation and stars to the Cosmic Microwave Background (CMB) and Large Scale Structure (LSS)~\cite{Ciarcelluti:2010zz,Ciarcelluti:2004ik,Ciarcelluti:2004ip}.

In this work, we study MTH cosmology and focus on the CMB and LSS signals. 
The Higgs portal coupling between the SM and twin particles has decoupled before the time relevant to the signals, and the two sectors only connect to each other through gravity. 
Before the twin protons and electrons recombine into neutral atoms, the twin baryons and photons are in kinetic equilibrium and experience the twin baryon acoustic oscillations (BAO). 
The twin BAO process suppresses the matter density perturbations at scale $k \gtrsim 0.1~h \,\rm{Mpc}^{-1}$ and introduces an oscillatory pattern in the matter power spectrum as for the standard BAO process. 
Encouragingly, the $k$ scale can be tested using current weak gravitational lensing surveys (hereafter called weak lensing). In particular, we focus on weak lensing measurements from the Dark Energy Survey (DES)~\cite{DES:2017myr, DES:2021wwk}, which are mainly sensitive to the range $0.1 \lesssim k \lesssim 10~h\,\rm{Mpc}^{-1}$.

Weak lensing can be used to study the LSS of the late universe by analyzing the shape distortions of a large number of galaxies due to the foreground matter field along the line of sight. The full set of weak lensing measurements called `\texttt{3x2pt}' contains 
three sets of two-point correlations with the angular separation $\theta$ of galaxy pairs:
galaxy clustering $w(\theta)$ (position-position), 
galaxy-galaxy lensing $\gamma_t(\theta)$ (position-shape), and 
cosmic shear $\xi_\pm(\theta)$ (shape-shape). 
The quantity $w(\theta)$ measures the distribution of angular separation of foreground lens galaxies compared with random distribution, 
$\gamma_t(\theta)$ measures the correlation 
between the distribution of foreground lens galaxies and the shape distortion of background source galaxies at $\theta$, and  
$\xi_\pm(\theta)$ measures the correlations between the shape distortion of background source galaxies due to the foreground LSS.
Compared with the other two correlators, cosmic shear $\xi_\pm(\theta)$ directly measures the matter-matter power spectrum and 
can be used to constrain the cosmological parameters, especially for the $S_8$ amplitude defined as $S_8=\sigma_8\sqrt{\Omega_m/0.3}$~\cite{DES:2017myr}. Here 
$\sigma_8$ is the mass dispersion on a scale around $8~h^{-1}\rm{Mpc}$ 
and $\Omega_m$ is the total matter abundance. 
Since the twin BAO modifies the $S_8$ compared to the $\Lambda$CDM model, we adopt cosmic shear measurement for our likelihood to constrain the MTH parameters.   
Several ongoing surveys regarding cosmic shears, such as Dark Energy Survey (DES)~\cite{DES:2017myr, DES:2021wwk}, 
Kilo-Degree Survey (KiDS)~\cite{vanUitert:2017ieu, Heymans:2020gsg}, 
Subaru Hyper Suprime-Cam (HSC)~\cite{HSC:2018mrq,Hamana:2019etx}, 
have robustly constrained $S_8$. 
There are also future experiments scheduled to operate in the near future, such as Vera C. Rubin Observatory~\cite{LSST:2008ijt}, 
Euclid~\cite{EUCLID:2011zbd}, 
the Nancy Grace Roman Space Telescope~\cite{Green:2012mj}, 
the Wide Filed Survey Telescope (WFST)~\cite{Lou:2016,Lei:2023adp}, 
and China Space Station Telescope (CSST)~\cite{Zhan:2011,Zhan:2021,Gong:2019yxt}. These surveys will further improve the sensitivity to the MTH signals.

The existence of twin particles may not only generate unique cosmological signals but also be responsible for tensions in the current cosmological data. Although the $\Lambda$CDM model provides excellent descriptions of most cosmological and astrophysical observations, emerging tensions with measured values of $H_0$~\cite{DiValentino:2021izs} and the $S_8$ amplitude~\cite{MacCrann:2014wfa} suggest that the model may not be complete. The Hubble tension arises from the discrepancy between the $\Lambda$CDM $H_0$ value favored by the Planck CMB data~\cite{planck} and the value obtained by the SH0ES collaboration~\cite{Riess_2021}, as well as the local measurement of different distance ladders~\cite{DiValentino:2021izs}, with a near 5$\sigma$ statistical deviation. Independently, though weaker, evidence for the Hubble
tension also comes from the latest multi-messenger analysis of GW170817/GRB~170817A~\cite{Wang:2022msf}.  
On the other hand, from the $\Lambda$CDM fit of the Planck data~\cite{planck_s8}, 
the derived $S_8$ is larger than almost all the low-redshift measurements from weak lensing and galaxy cluster surveys, leading to a $2-3\sigma$ tension. Specifically, the datasets from KV450, KiDS-450, DESY1, and HSC-DR1 have all shown consistent results in this regard (see the summary plots in~\cite{Hildebrandt_2020}).  Although the $S_8$ tension is currently less significant than the $H_0$ tension, it may still imply the unexpected cosmological evolution\cite{Krishnan:2020vaf,Adil:2023jtu} and the existence of new physics beyond the $\Lambda$CDM model\footnote{It has been argued \cite{Ade:2013lmv, vonderLinden:2014haa, Umetsu:2020wlf, Blanchard:2021dwr,Nunes:2021ipq} that this tension could be the result of systematics uncertainties, for instance, in mass bias. However, in this work, we assume that the tension is an indication of new physics.}. As we anticipate future large-scale structure surveys to reduce uncertainties in the power spectrum measurement, a similar central value of $S_8$ obtained from these surveys would call for a simultaneous consideration of both the $H_0$ and $S_8$ tensions and identification of the cosmological model that addresses both.

One possibility for addressing both the $H_0$ and $S_8$ tensions is to introduce an interacting dark matter (DM) with dark radiation (DR) in addition to the $\Lambda$CDM model~\cite{Lesgourgues:2015wza,Chacko:2016kgg,Foot:2016wvj,Buen-Abad:2017gxg,Chacko:2018vss,Dessert:2018khu,Bansal:2021dfh,Joseph:2022jsf,Bansal:2022qbi,Buen-Abad:2022kgf}. 
Several models, such as those incorporating early dark energy \cite{Poulin:2018cxd} or featuring an additional effective number of neutrino species ($\Delta N_{\rm eff}$), 
have been proposed to explain the larger value of $H_0$ inferred from Planck data (see, e.g., reviews by~\cite{DiValentino:2021izs,Schoneberg:2021qvd}). However, these models typically exacerbate the $S_8$ tension if an additional source of energy density exists around the time of matter-radiation equilibrium~\cite{Anchordoqui:2021gji}. 
On the other hand, models featuring interactions between DM and DR have the potential to simultaneously increase the value of $H_0$ and alleviate the $S_8$ tension by leveraging acoustic oscillations resulting from DM-DR scattering~\cite{Munoz:2020mue,Bohr:2020yoe}. 
One such example is the MTH model, where twin protons and twin electrons act as interacting DM before twin recombination, and the resulting twin BAO process further mitigates the $S_8$ tension~\cite{Bansal:2021dfh}. 
In Ref.~\cite{Bansal:2021dfh}, the authors used the $S_8$ value from the Planck~SZ~(2013) analysis~\cite{Ade:2013lmv} to demonstrate that the MTH model can alleviate both the $H_0$ and $S_8$ tensions. 
Despite the Planck~SZ~(2015) report~\cite{Planck:2015lwi} indicating that the 2013 analysis underestimates the uncertainty in the $S_8$ measurement, 
the 2013 result can still be used to test the MTH model's ability to reconcile these significant tensions. 
Furthermore, since the central value of the $S_8$ from the Planck~SZ~(2013) is in close agreement with the KV450 result, 
future LSS measurements may also converge to the Planck~SZ~(2013) value with reduced uncertainty. Unlike the analysis in~\cite{Bansal:2021dfh}, 
which only utilizes the KV450 dataset  up to $k_{\rm max}=0.3\,h$Mpc$^{-1}$ to constrain the LSS, 
our study incorporates the current DES data, extending the range to $k_{\rm max}=10\,h$Mpc$^{-1}$.\footnote{
To reduce the impact of non-linear baryonic effects, the DES analysis in~\cite{DES:2021bvc} masked out the lensing two point correlation function within certain galaxy separation angles. Our likelihood study shows that the resulting DES data is only sensitive to the change of matter power spectrum within $k\lesssim10\,h$Mpc$^{-1}$.
}  
Our results demonstrate that while satisfying the DES constraint even at higher $k$ modes, 
the MTH model effectively relaxes the $H_0$ and large $S_8$ tensions to a degree comparable to the Planck~SZ~(2013) result around $k\sim 0.1\,h$Mpc$^{-1}$.

This paper is organized as follows. In Sec.~\ref{sec:model}, we briefly introduce the MTH model configuration and its cosmological history. We then describe the measurement and MTH model prediction of the two-point correlation function of cosmic shear in Sec.~\ref{sec:shear}. In Sec.~\ref{sec:mcmc}, we summarize our numerical methodology for performing a high-dimensional Markov Chain Monte Carlo (MCMC) scan. Our present constraints, which include CMB, BAO, DES~Y3, SH0ES, and Planck~SZ~(2013), are presented and discussed in Sec.~\ref{sec:result}. Furthermore, we estimate the sensitivity of cosmic shear for the CSST survey in comparison to DES Y3 data within the MTH parameter space. In Sec.~\ref{sec:N_body}, we justify the use of \texttt{HMCode} for non-linear correction by comparing a few results from $N$-body simulations. We conclude in Sec.~\ref{sec:sum}.

\section{The Mirror Twin Higgs model} 
\label{sec:model}

In the MTH model, each SM particle has its counterpart in the twin sector, including the twin radiation (twin photons and neutrinos) and the twin matter (twin baryons and leptons). The twin sector has the same gauge symmetries and dimensionless couplings as the ordinary world. However, the mirror ($Z_2$) symmetry between the two sectors is softly broken, which leads to different vacuum expectation values (VEVs) in the electroweak symmetry breaking. Although the interactions from the Higgs or photon mixings between two sectors may leave signatures in the DM direct~\cite{Foot_2014,Chacko:2021vin,Zu:2020idx} and indirect~\cite{Curtin:2019lhm,Curtin:2019ngc} detections, 
the experimentally allowed portal couplings are too weak to alter the cosmological evolution. Thus, we neglect the non-gravitational interactions between the two sectors in our cosmological study.
We follow Ref.~\cite{Chacko:2018vss,Bansal:2021dfh} to model the MTH cosmology by three independent parameters (where $\hat{X}$ refers to the twin copy of the SM $X$ particle or energy scale):  
\begin{itemize}
    \item $\Delta \hat N=\Delta N_{\hat\nu}+\Delta N_{\hat\gamma}$ from the twin neutrinos and photons with $\Delta N_{\hat\nu}/\Delta N_{\hat\gamma}=4.4/3$. The twin and SM temperature ratio can be inferred by $\hat T/T=(\Delta\hat N/7.4)^{1/4}$.
    \item Energy density ratio $\hat{r}\equiv\frac{\Omega_{\hat b}}{\Omega_{\rm cdm}}$ today. The twin baryon density $\Omega_{\hat b}$ is the sum of the twin hydrogen and helium energy densities.
    \item VEV ratio $\hat{v}/v$ between the twin and SM electroweak symmetry breaking.     
\end{itemize}
If the temperature of the twin sector is the same as that in SM, 
the twin photons and neutrinos as DR predict $\Delta\hat N\sim 7$,  
which has been excluded by Planck CMB observation, namely $\Delta N_{\rm eff} < 0.3$ in $95\%$ credible region~\cite{planck}. 
To reduce the $\Delta N_{\rm eff}$, an idea is to cool down the twin sector. 
For example, introducing an asymmetric post-inflationary reheating mechanism 
to be more efficient in the SM than the twin sector~\cite{Chacko_2017,Craig_2017,Beauchesne_2021}. 

Since the interactions between twin particles mimic their partners in the SM sector, 
the thermal history of twin baryons and leptons is similar to the SM sector. Some important cosmological events therefore also take place in the twin sectors, including the Big Bang Nucleosynthesis (BBN) that determines the ratio between helium and hydrogen densities, and the recombination process when the ionized hydrogen and helium combine into neutral atoms. With given MTH parameters defined above, we can predict the amount of twin helium vs. hydrogen fraction and the visibility function of the twin photons that are important to determining the cosmological signatures.


In this work, we use a modified version of public Boltzmann code \texttt{CLASS}~\cite{Blas:2011rf} made by the authors in~\cite{Bansal:2021dfh} to calculate the CMB and matter power spectra with the presence of MTH particles. The modified code calculates the mass fraction of twin helium $Y_{\rm p}(^4{\rm He})=\rho_{\hat{\rm H}{\rm e}}/(\rho_{\hat{\rm H}{\rm e}}+\rho_{\hat{\rm H}})$ and obtain the slip term that couples the velocity divergence perturbations between the twin baryon and photon. The code than pass the calculated numbers to the ETHOS~\cite{Cyr_Racine_2016} module to obtain the evaluation of the twin hydrogen/helium/photon energy density perturbations before and after the recombination. We treat twin neutrinos as free streaming radiation in the study. More details about the MTH cosmology have been discussed in~\cite{Bansal:2021dfh}.  

\begin{figure}[t!]
    \centering
    \includegraphics[scale = 0.37]{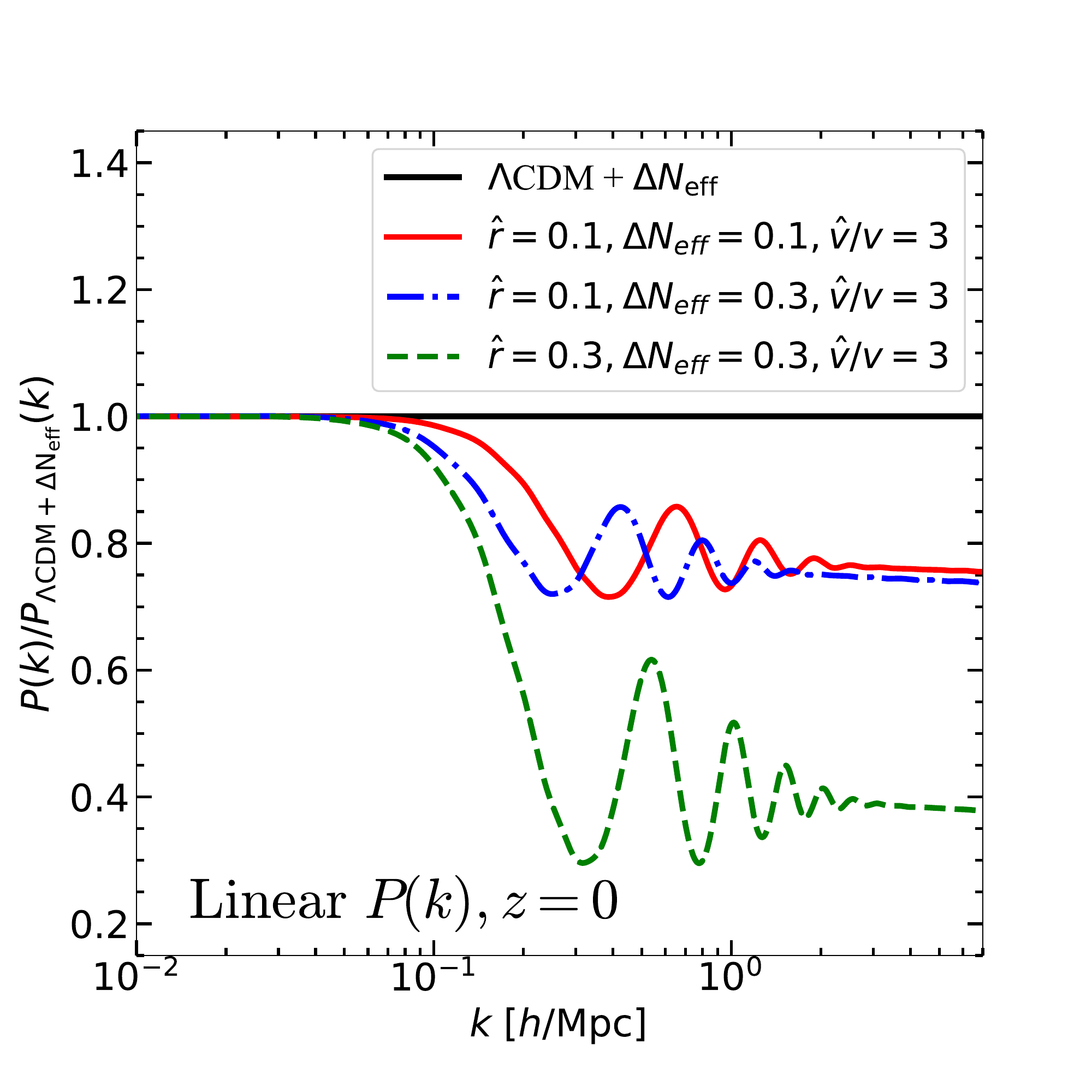}
    \includegraphics[scale = 0.37]{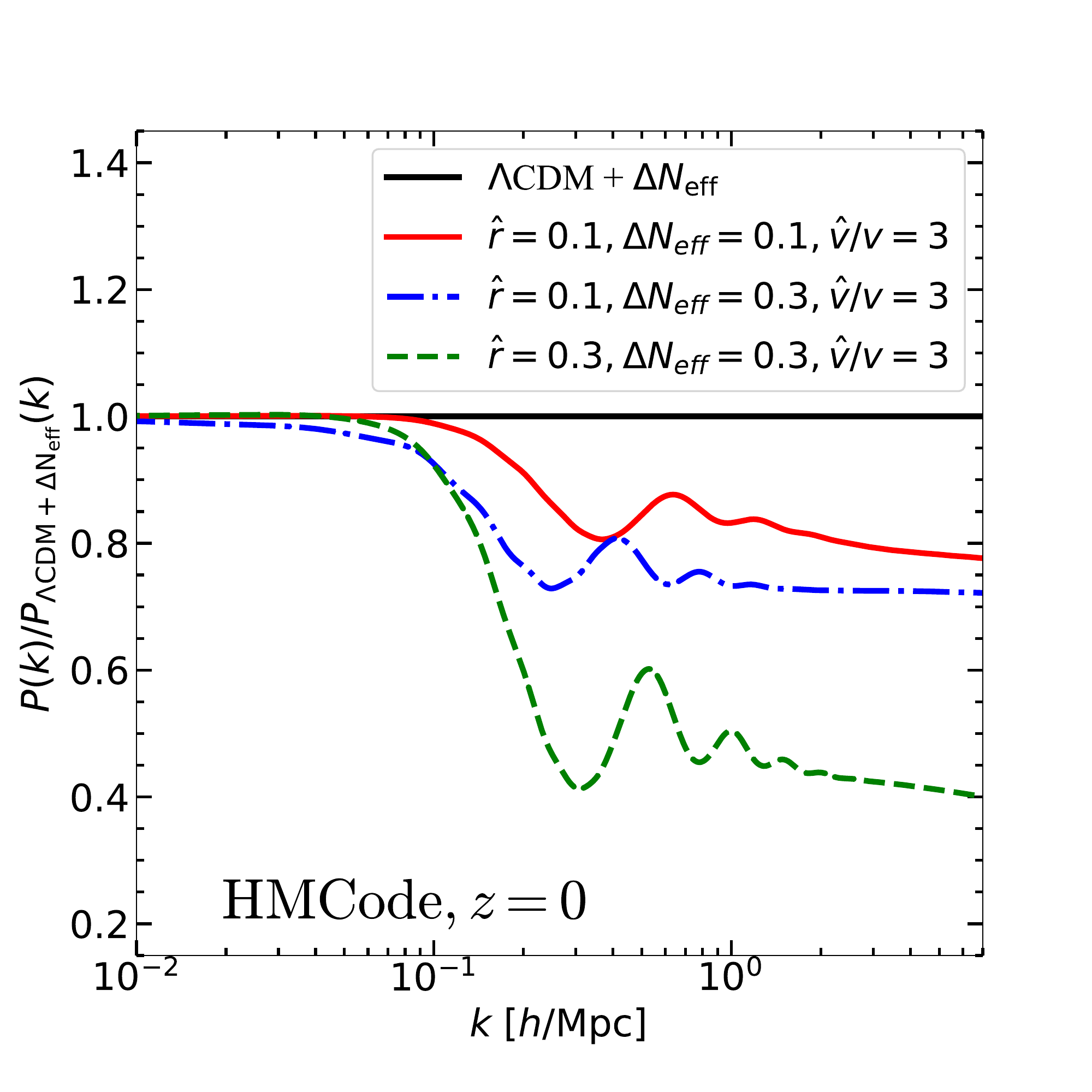}
    \caption{The linear (left panel) and non-linear (right panel) twin Higgs matter power spectrum for various twin parameters  
     at $z=0$. Their cosmological parameters used for calculation are $h=0.676$, $\Omega_b h^2=0.0228$, $\Omega_{\rm cdm} h^2=0.121$, 
    $\rm{ln}(10^{10}A_s)=3.05$, $n_s=0.976$, and $ \tau_{\rm reio}=0.061$. 
    }
    \label{Fig:p_ratio}
\end{figure}

The CMB and matter power spectra in MTH cosmology differ from those in the $\Lambda$CDM because of the presence of twin radiation and BAO~\cite{Bansal:2021dfh}. For the CMB TT and EE power spectra, the twin BAO suppresses the metric perturbations and enhances (decreases) the expansion (contraction) mode of temperature perturbations. In addition, scattering DR from the twin photon also changes the amount of phase shift for the oscillation peaks in the TT and EE spectra compared to the free-streaming DR scenarios. Both effects leave visible signals in the Planck measurement. 

For the matter power spectrum, the presence of twin BAO suppresses the matter density perturbations at a small scale. Moreover, it introduces an oscillation in the power spectrum in addition to the SM BAO. In the left panel of Fig.~\ref{Fig:p_ratio}, we present three benchmark linear matter power spectra with the same cosmological inputs $h=0.676$, 
$\Omega_b h^2=0.0228$, $\Omega_{\rm cdm} h^2=0.121$, $\rm{ln}(10^{10}A_s)=3.05$, $n_s=0.976$, $ \tau_{\rm reio}=0.061 $. At small scale, $k \gtrsim 0.1~h\rm{Mpc}^{-1}$ that enters the horizon before the twin recombination epoch, 
the linear matter power spectrum deviates from that of $\Lambda\rm{CDM}+\Delta N_{\rm eff}$ with an oscillation and a suppression. 
Thus, the matter power spectrum in the scale $k \gtrsim 0.1~h\rm{Mpc}^{-1}$ is one of the best regions to test the MTH model. 
Coincidentally, this scale structure is testable by using weak lensing observations. In this work, we probe the matter distribution in this scale based on the DES cosmic shear data. 

The linear evolution of density perturbations assumes 
the fractional density fluctuations $\delta(\rm{k}) \ll 1$ so that correction from $\mathcal{O}(\delta(\rm k)^2)$ terms is negligible. However, in the late universe, $\delta(\rm{k})$ can be comparable to or even greater than unity at $k =\mathcal{O}(0.1)~h {\rm Mpc}^{-1}$ and above, and the linear calculation no longer applies. Therefore, like computation performed in IDM-DR model~\cite{Schaeffer_2021,Munoz:2020mue}, 
we should, in principle, run $N$-body simulations to incorporate the non-linear corrections to the matter power spectrum up to $k\approx 10~h {\rm Mpc}^{-1}$ for the weak lensing analysis. However, the $N$-body simulation is much more CPU-time expansive, and having an event-by-event non-linear correction from the $N$-body simulation is unrealistic for the MCMC study. 

In this work, we take the following approach to estimate the non-linear correction. First, as discussed later in Sec.~\ref{sec:N_body}, we run $N$-body simulations for two benchmarks MTH models and show that the resulting $P(k)$ correction is similar to the one obtained from the non-linear correction code, \texttt{HMCode}~\cite{Mead_2016}. We then use the \texttt{HMCode} in the MCMC study for the non-linear correction at different redshift. In the right panel of Fig.~\ref{Fig:p_ratio}, we show the power spectrum with non-linear correction  
computed by \texttt{HMCode} code. 
Compared with the linear power spectrum (the left panel), 
the non-linear effects smear the twin BAO structure at $z=0$, as discussed in~\cite{Schaeffer_2021} for a more general setup of DAO models.

\section{Cosmic shear} 
\label{sec:shear}

In this section, we describe the calculation of the two-point correlation function of the cosmic shear from the observational data and explain how we obtain the covariance matrix that models
the uncertainties of the shear measurement. We also show an example of the shear spectrum for the MTH cosmology.

\subsection{Cosmic shear measurement}
The DES Y3 data measures the shapes of over $10^8$ source galaxies 
covering a footprint ${\rm A_{eff}} = 4143~\rm{deg}^2$. 
Following the DES Y3 analysis, the shape catalog 
\texttt{METACALIBRATION}\footnote{\url{https://desdr-server.ncsa.illinois.edu/despublic/y3a2_files/y3kp_cats/}}
is divided by 4 redshift tomographic bins in the range of $0<z<3$~\cite{DES:2021bvc}. 
In this work, we  calculate the two-point correlation function of  cosmic shear in the same manner.

Given a pair of galaxy images with angular  separation $\theta$ in any two redshift bins ($i$ and $j$), we can express the shear-shear correlation function in terms of ellipticities that include the tangential component $\epsilon_t$ and the cross component $\epsilon_\times$,
\begin{equation}
    \xi_\pm^{i j}(\theta)=
    \langle \epsilon^i_t \epsilon^j_t \rangle \pm 
    \langle \epsilon^i_\times \epsilon^j_\times \rangle\,.     
    \label{eq:xipmij}
\end{equation}
Here the brackets represent the average of all galaxies pairs. 
From the experimental observations, we can calculate the correlation function by summing 
all galaxy pairs (in the $i$ and $j$ redshifts) with indices $a$ and  $b$~\cite{DES:2021vln,DES:2017qwj}, 
\begin{equation}
\xi_{\pm}^{i j}(\theta)=
\frac{\sum_{a b} w_a w_b\left(\hat{e}_{t, a}^i \hat{e}_{t, b}^j \pm \hat{e}_{\times, a}^i \hat{e}_{\times, b}^j\right)}
{\sum_{a b} w_a w_b R_a R_b},
\label{eq:exp_xi}
\end{equation}
where $w_a$ and $w_b$ are the weight of each galaxy pairs in $i$ and $j$ redshift bin, respectively. 
Notice that we modify the ellipticity expression in Eq.~\eqref{eq:exp_xi}  by subtracting the residual mean shear, $\hat{e}_{k}^i = \epsilon_k^i - \langle \epsilon_k \rangle^i$. The response correction matrix $R$ comes from the linear order expansion of the ellipticities from a noisy and biased measurement ($\hat{e}$) with respect to the gravitational shear ($\xi_{\pm}$) that we want to extract~\cite{DES:2020ekd}. $R$ includes the measured shear response matrix $R_\gamma$ and the shear selection bias matrix $R_S$. 
We can explicitly write down the response correction as $R_{[k,l]}=R^{[k,l]}_{\gamma}+R^{[k,l]}_{S}$ for the $[k,l]$ element of the matrices, where $k,l=1,2$ that corresponds to the two shear degrees of freedom $t$ and $\times$. Following the analysis in Ref.~\cite{DES:2017qwj}, we take an average of each shear components $R_{a (b)}=(R_{[1,1]}+R_{[2,2]})/2$ for a galaxy $a (b)$ when calculating Eq.~(\ref{eq:exp_xi}). We take the ellipticities $\hat{e}_{k}^i$ and the response correction matrix $R = R_\gamma +R_S$ from 
shape catalog \texttt{METACALIBRATION}. 
The cosmic shear data vector $\xi_\pm^{ij}$ can be obtained by using public code {\texttt{TreeCorr}}~\cite{Jarvis:2003wq}.  All the auto correlations and cross correlations of the four redshift bins are calculated using 20 $\theta$-bins arranging logarithmically from $2.5$ to $250$ arcmin. 

\subsection{Covariance matrix}
\label{covmat}
We model the statistical uncertainties of the $\xi_\pm$ measurement by hiring a covariance matrix $\mathbf{C}$ 
which can be calculated with public code \texttt{CosmoCov}~\cite{Fang:2020vhc,Krause:2016jvl}. Note that the matrix $\mathbf{C}$ strongly depends on the per-component shape dispersion $\sigma_e$, 
galaxies' redshift distribution (galaxy number densities), and cosmological parameters. Therefore, we should embed an unfixed $\mathbf{C}$ when performing a cosmological MCMC scan. However, such a scan is numerically challenging, and we instead take an approximation following 
the same analyses in the cosmic shear literature~\cite{DES:2021bvc}. First, we directly take the per-component shape dispersion $\sigma_e$ and galaxies effective number densities $n_{\rm eff}$ 
used in the covariance matrix from Table I of Ref.~\cite{DES:2021bvc} in the analysis. 
In order to shorten the scanning time, we use a set of fiducial parameters for the initial MCMC chains, 
namely \{$\Omega_m = 0.3$, $\Omega_\Lambda=0.7$, $\sigma_8=0.82355$, $n_s=0.97$, $\Omega_b=0.048$, $h = 0.69$\}. 
After the initial scans converge, we recompute the covariance matrix 
by using the best-fit parameters \{$\Omega_m = 0.306$, $\Omega_\Lambda=0.694$, 
$\sigma_8=0.802$, $n_s=0.974$, $\Omega_b=0.022$, $h = 0.686$\} and fix them for the later scans. 

\subsection{DM model prediction}

The MTH theoretical prediction of the cosmic shear for two redshift bins ($i$ and $j$) is expressed as
\begin{equation}
\xi^{ij}_\pm(\theta) = \frac{1}{2\pi}{\int_{0}^{\infty}C_{\rm tot}^{i j}(\ell)J_{0/4}(\ell\cdot \theta) \ell d\ell},  
\end{equation}
where $\ell$ is the angular wave number.  
The Bessel function is the zeroth-order $J_{0}$ for $\xi_+$, but fourth-order $J_{4}$ for $\xi_-$. 
Under Limber approximation~\cite{Limber:1953,LoVerde:2008re}, 
we can derive the 2D shear-shear power spectrum $C_{GG}^{i j}(\ell)$ between redshift bin $i$ and $j$ 
in terms of the 3D matter power spectrum $P_\delta$, 
\begin{equation}
\label{eq:cl_gg}
C_{GG}^{i j}(\ell)=\int_0^{\chi_{\rm z_{max}} } d \chi \frac{W^i(\chi) W^j(\chi)}{\chi^2} P_\delta\left(k = \frac{\ell+1 / 2}{\chi}, z(\chi)\right),
\end{equation}
where $\chi$ is the comoving distance.
We shall also consider contribution from the intrinsic alignment (IA) given in Appendix~\ref{sec:IA}, 
thus the total 2D spectrum is
$C_{\rm tot}^{ij}(\ell) = C_{GG}^{ij}(\ell)+C_{GI}^{ij}(\ell)+C_{II}^{ij}(\ell)$. 
To save computation time, we only calculate  
the non-linear power spectrum ${\rm P_\delta}$ up to $k = 10~h\rm{Mpc}^{-1}$ 
with the \texttt{HMCode} in \texttt{CLASS}. 
The lensing efficiency kernel $W^i(\chi)$ in the redshift bin $i$ is given by
\begin{equation}
W^i(\chi)=\frac{3 H_0^2 \Omega_{\mathrm{m}}}{2 c^2} \frac{\chi}{a(\chi)} \int_\chi^{\chi_{\rm z_{max}}} d \chi^{\prime} n^i\left(\chi^{\prime}\right)
\frac{\chi^{\prime}-\chi}{\chi^{\prime}},
\end{equation}
where $n^i(z)$ is the normalized galaxies redshift distribution in the $i$-th redshift bin, and 
we have $\int n(\chi)d\chi=\int n(z)dz=1$.
The scale factor is $a(\chi)$, 
and $c$ is the speed of light. It needs to note that in our analysis we have neglected the multiplicative bias and photometric redshift calibration uncertainties since both of them only affect the final results with a tiny deviation $\sim O(0.001)$~\cite{DES:2021bvc}.

\begin{figure}[t!]
    \centering
    \includegraphics[scale = 0.37]{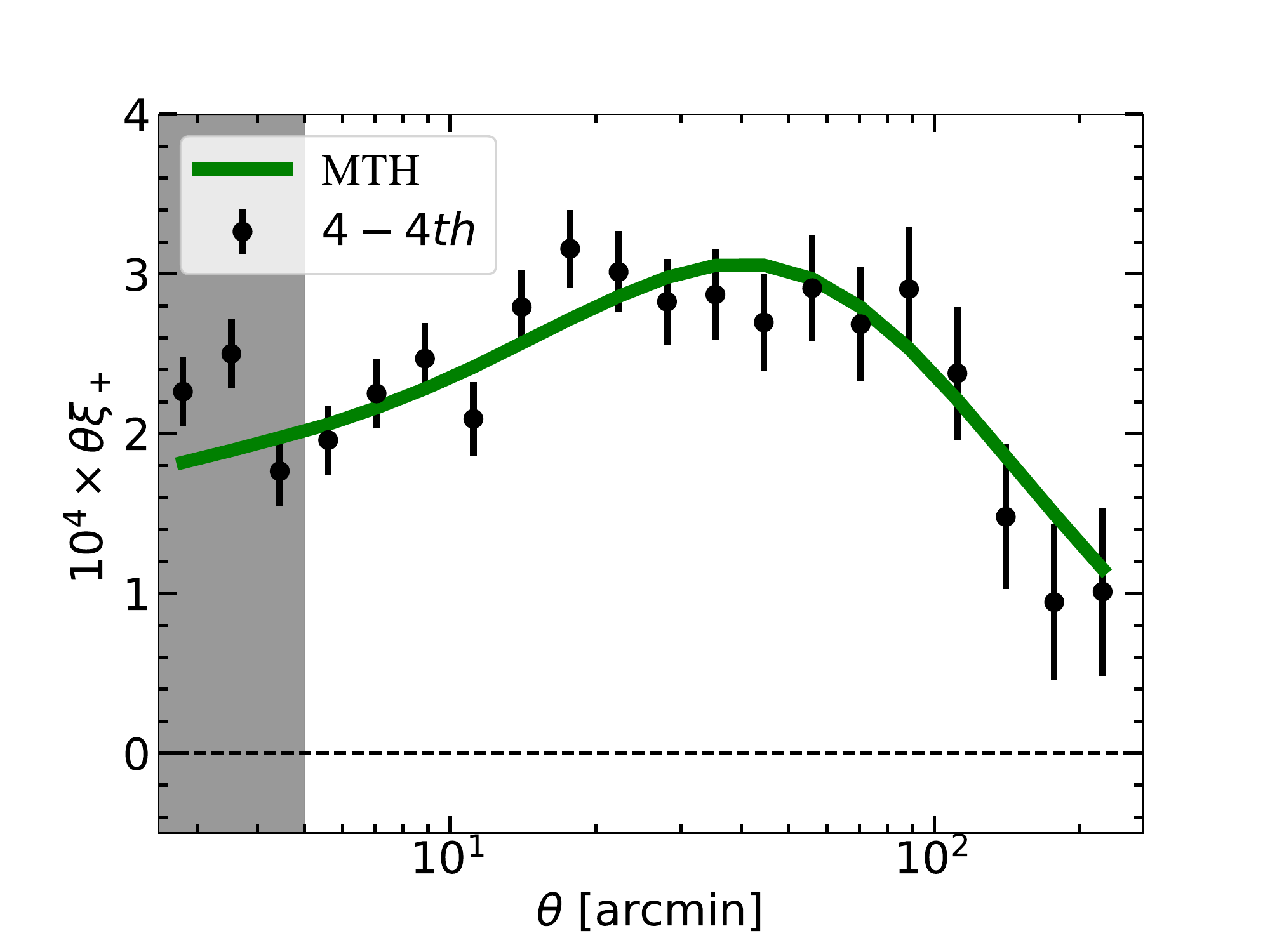}
    \includegraphics[scale = 0.37]{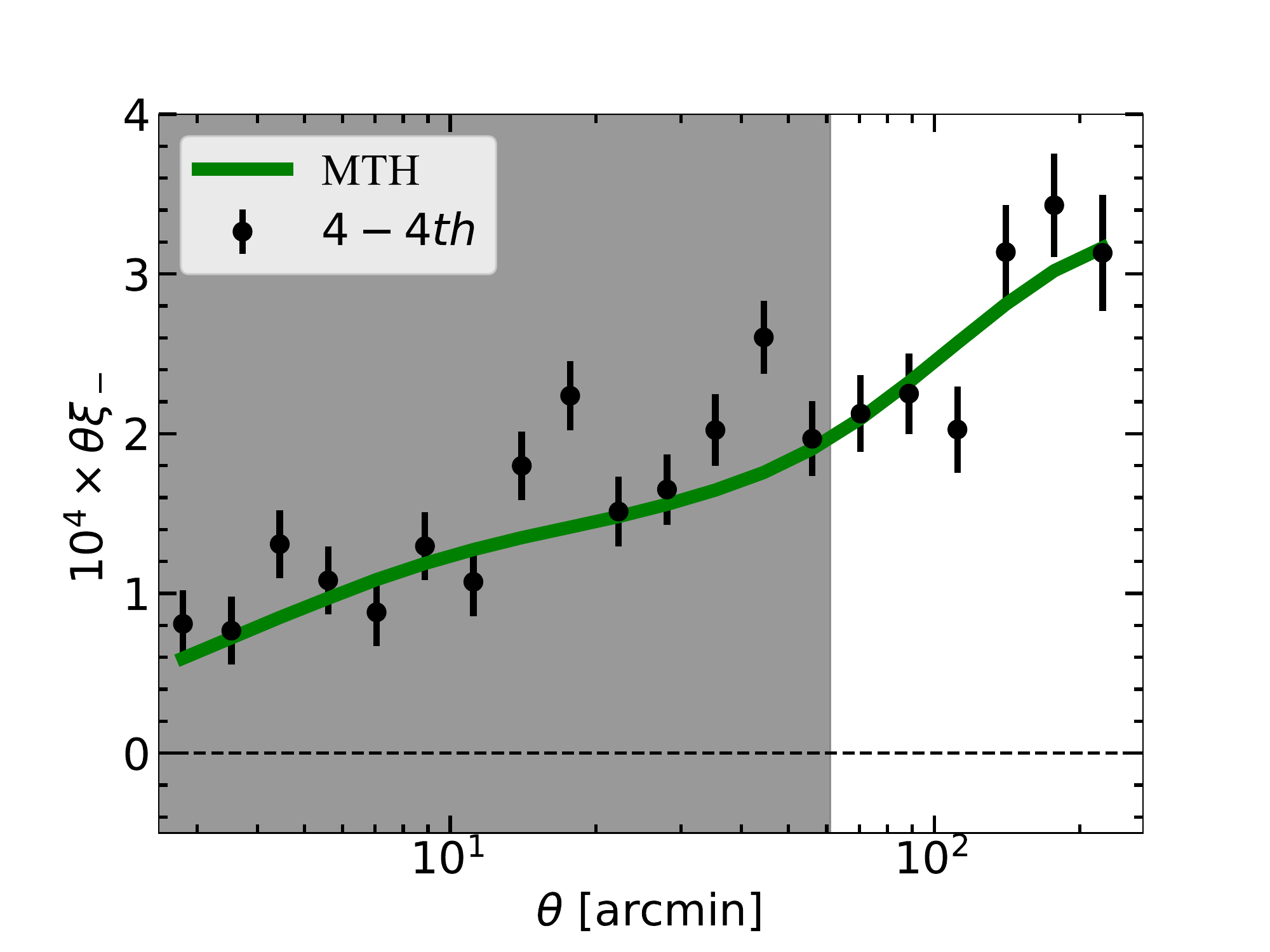}
    \caption{The auto two-point correlation function of cosmic shear for 4-4$th$ redshift bin. 
    The data points correspond to DES Y3 cosmic shear measurement. 
    The diagonal term of the covariance matrix gives the error bars of data points. 
    The green solid lines are computed by \texttt{HMCode}. 
    The gray region is masked to avoid the small-scale uncertainties as applied in Ref.~\cite{DES:2021bvc}.}
    \label{xi_pm44}
\end{figure}

In Fig.~\ref{xi_pm44}, we demonstrate $\xi^{44}_+$ (left panel) and $\xi^{44}_-$ (right panel) as an example, 
because the 4-4$th$ redshift bin presents the most foreground lensing effect. 
For the results of the rest redshift bins, we refer them to  
Fig.~\ref{xi_pm_tot} in Appendix.~\ref{sec:appendix_fig}. 
Here, the black data points present the DES Y3 cosmic shear measurement, and the error bars denote the diagonal term of the covariance matrix. To reduce the impact of non-linear baryonic effects, we exclude data below certain $\theta$ (gray) as recommended in Ref.~\cite{DES:2021bvc}, which is based on hydrodynamic simulations.
As a comparison, we compute the best-fit MTH prediction (green solid line) based on the parameter configuration: 
\{$\Omega_b h^2 = 0.023$, $\Omega_{\rm cdm} h^2 = 0.121$, $\theta_s = 1.042$, $\rm{ln}(10^{10}A_s)=3.034$, $n_s = 0.975$, $\tau_{\rm reio} = 0.051$, $\hat{r} = 0.067$, $\hat{v}/v = 2.725$, $\Delta \hat N = 0.168$\}.

\section{Methodology} 
\label{sec:mcmc}

In this section, we briefly discuss the likelihoods and priors used in the MCMC scan 
of the MTH and cosmological parameter space.
We employ Bayes's theorem to find the posterior probability density function, 
\begin{equation}
    {\rm Posterior}\propto {\rm Likelihood} \times {\rm Prior}. 
\end{equation}
The normalization of the posterior called ``evidence'' is a quantity relevant to the model comparison. 
Our results are presented by marginalizing the posterior densities of unwanted parameters 
and showing the $68\%$ and $95\%$ credible regions.

\begin{table}
    \centering
\begin{tabular}{c|c|c}
\hline
\hline
    Parameter & Prior distribution & Prior range \\
\hline
\hline
\multicolumn{3}{|c|}{\textbf{Cosmology}}\\
\hline
\hline
    $\Omega_b h^2$ & Flat & [0.022, 0.023] \\
    $\Omega_{\rm cdm} h^2$ & Flat & [0.112, 0.128] \\
    $100 \cdot \theta_s$ & Flat & [1.039, 1.043] \\
    $\ln\left(A_s\times 10^{10}\right)$ & Flat & [2.955, 3.135] \\
    $n_s$ & Flat & [0.941, 0.991] \\
    $\tau_{\rm reio}$ & Flat & [$10^{-2}$, 0.7] \\
\hline
\hline
\multicolumn{3}{|c|}{\textbf{Mirror twin Higgs}}\\
\hline
\hline    
    $\hat{r}$ & Flat & [$10^{-3}$, 1] \\
    $\hat{v}/v$ & Flat & [2, 15] \\
    $\Delta\hat N$ & Flat & [$10^{-3}$, 1] \\
\hline
\hline
\multicolumn{3}{|c|}{\textbf{Intrinsic alignment}}\\
\hline
\hline
    $A_{\rm IA}$ & Flat & [-6, 6] \\
    $\eta$ & Flat & [-6, 6] \\
\hline
\hline
\end{tabular}
    \caption{All the input parameters used in our scan. 
    Three types of parameters are grouped as cosmological, MTH, and intrinsic alignment parameters.}
\label{tab:mcmc_prior}
\end{table}

To explore the MTH model parameter space, 
we conduct a global scan by using a public cosmological MCMC package 
\texttt{Monte Python}~\cite{Brinckmann:2018cvx,Audren:2012wb}, 
and it interfaces with \texttt{CLASS} to compute the cosmological observables. 
In Table~\ref{tab:mcmc_prior}, we show the prior ranges used in our numerical scan, including $\Lambda$CDM parameters as defined in~\cite{planck}
and MTH parameters: 
the fraction of the twin baryon to DM energy density $\hat{r}$, 
the ratio of vacuum expectation value between the twin and SM sectors $\hat{v}/v$, and 
the extra effective number of neutrinos from the twin sector $\Delta \hat{N}$. 

We apply five likelihood sets in the MCMC study. For setting existing bounds on the MTH model, we consider three datasets
\begin{enumerate}
    \item[(i)] The cosmic shear likelihood based on the DES Y3 \texttt{METACALIBRATION} shape catalog, which described in Sec.~\ref{sec:shear}. 
    \item[(ii)] The CMB likelihoods are calculated based on Planck 2018 Legacy~\cite{Planck:2019nip}, 
    including high-$\ell$ power spectra (\texttt{TT}, \texttt{TE}, and \texttt{EE}), 
    low-$\ell$ power spectrum (\texttt{TT} and \texttt{EE}), and Planck lensing power spectrum (\texttt{lensing}).
    \item[(iii)] The BAO likelihood contains the BOSS DR12 dataset measurements at $z=0.106$, $z=0.15$ and $z=0.2-0.75$ ~\cite{Beutler_2011,Ross_2015,Alam_2017}.
\end{enumerate}  
When studying how MTH can relax the $H_0$ and $S_8$ tensions, we further include two datasets one by one
\begin{enumerate}
\item[(iv)] The SH0ES likelihood is also a Gaussian distribution with the measurement~\cite{Riess:2021jrx}
\begin{equation}
       H_0=73.04 \pm 1.04 \mathrm{~km} \mathrm{~s}^{-1} \mathrm{Mpc}^{-1}. \nonumber
    \end{equation}

\item[(v)] The Planck SZ (2013) likelihood\footnote{As mentioned in the Introduction, we use the Planck~SZ~(2013) data only as a demonstration of a potential signal scenario, since the $S_8$ tension significance from the later Planck~SZ~(2015) analysis~\cite{Planck:2015lwi} is reduced.} 
is described by a Gaussian distribution 
    with the measurement~\cite{Ade:2013lmv} 
    \begin{equation}
        S_8^{\mathrm{SZ}} \equiv \sigma_8\left(\Omega_{\mathrm{m}} / 0.27\right)^{0.3}=0.782 \pm 0.010. \nonumber
    \end{equation}
\end{enumerate}
We would like to note that $S_8^{\mathrm{SZ}}$ here differs from the definition of $S_8$ used in this work. The $S_8^{\rm SZ}$ value is in $2-3\sigma$ tension to the $\Lambda$CDM fit of the CMB data. However, as mentioned in the introduction, the measurements of galaxy distribution from the SZ effect depend on a mass bias factor $(1 - b)$ that relates the observed SZ signal to the true mass of galaxy clusters. The Planck~SZ~(2013) report gives an $S_8$ measurement by fixing the mass bias to its central value from a numerical simulation, $(1-b)=0.8$. 
The Planck~SZ~(2015) report later allowed $(1-b)$ to vary with a Gaussian prior centered at $0.79$. The central value of the resulting $S^{\rm SZ}_8$ becomes smaller
but has a much larger uncertainty $S_8^{\rm SZ}=0.744\pm 0.034$~\cite{Planck:2015lwi} and less tension to CMB measurements. Nevertheless, since the central value of the $\sigma_8$ from the SZ~(2013) analysis ($\sigma_8=0.76$ for $\Omega_m=0.3$) is consistent with most of the low-redshift measurements~\cite{2022JHEAp..34...49A}, future LSS measurements may converge to the SZ~(2013) result with a better-determined bias factor. Therefore, including the SH0ES and the SZ~(2013) datasets help demonstrate the MTH model's capability to simultaneously solve the $H_0$ and $S_8$ problems when both tensions increase to $3-4\sigma$ in the future. 

In the MCMC study, we implement all the likelihood from the \texttt{Monte Python} except for the DES cosmic shear. We implement the cosmic shear likelihood function as
\begin{equation}
\ln \mathcal{L}_{\rm shear}=-\frac{1}{2} \sum_{i j}\left(D_{i}-T_{i}\right)
\left[\mathbf{C}^{-1}\right]_{i j}
\left(D_{j}-T_{j}\right), 
\label{eq:cslike}
\end{equation}
where $T$ and $D$ are the theoretically predicted and experimentally measured shear-shear correlation functions, respectively. 
We obtain the covariance matrix $\mathbf{C}$ from a procedure described in Sec.~\ref{covmat}. 
Note that we mask the small scale data to avoid the small scale uncertainties as applied in Ref.~\cite{DES:2021bvc}, 
because the baryon feedback effect at small scale (e.g. AGN feedback or star-forming feedback)
remains largely unknown.

We perform four MCMC scans step-by-step for the MTH model based on the likelihoods stated in the section: \texttt{Planck+BAO}, \texttt{Planck+BAO+DES~Y3}, \texttt{Planck+BAO+DES~Y3+SH0ES} and \texttt{Planck+BAO+DES~Y3+SZ (2013)+SH0ES}. 
As a comparison, four scans for $\Lambda$CDM based on the same likelihoods are also performed. 
In each scan, we collect 8 chains with approximately 0.5 million samples in total. 
The average acceptance rate is around $0.2$.

\section{Results}
\label{sec:result}
This section presents the MTH parameter space favored by four different combinations of datasets described in Sec.~\ref{sec:mcmc}:   
(i) by fitting the \texttt{Planck+BAO} and \texttt{Planck+BAO+DES Y3} data, we show the current credible region of the MTH parameters. 
(ii) by fitting the \texttt{Planck+BAO+DES~Y3+SH0ES} data, we show that the MTH model can accommodate a larger $H_0$ and smaller $S_8$ compared to the $\Lambda$CDM and $\Lambda$CDM+$\Delta N_{\rm eff}$ models. 
(iii) to further examine MTH model's ability to resolve the $S_8$ tension, we compare the MTH and $\Lambda$CDM models based on the likelihood 
\texttt{Planck+BAO+DES~Y3+SH0ES+SZ (2013)}. 
(iv) finally, to show the prospect of measuring the MTH parameters with the future CSST, we generate mock data using a benchmark MTH model and perform a fit to the \texttt{Planck+BAO+SH0ES+SZ~(2013)+CSST} likelihood.

\subsection{CMB, BAO and Weak Lensing} 
\label{sec:CMB_BAO_lensing}

\begin{figure}[t!]
    \centering
    \includegraphics[scale = 0.35]{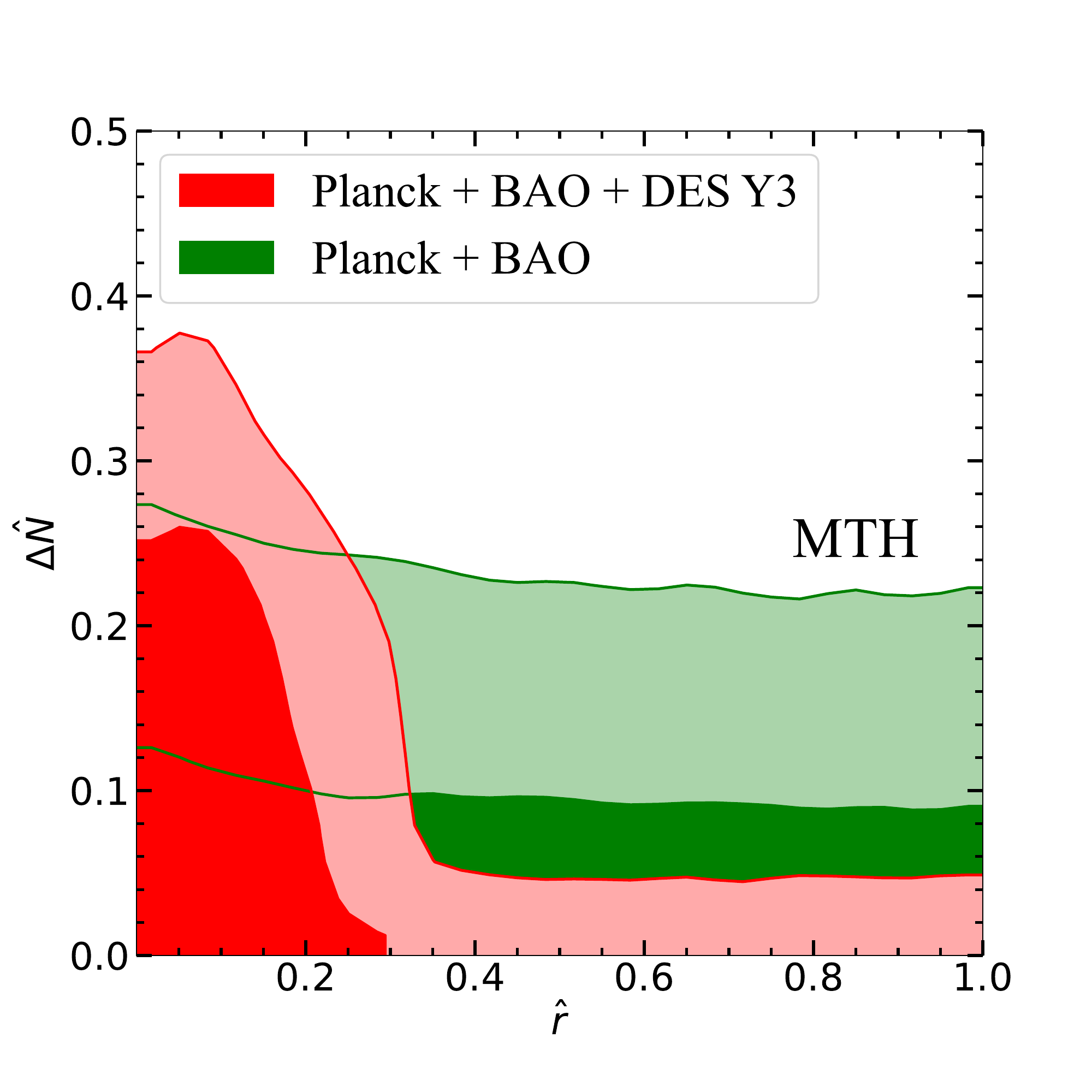}
    \includegraphics[scale = 0.35]{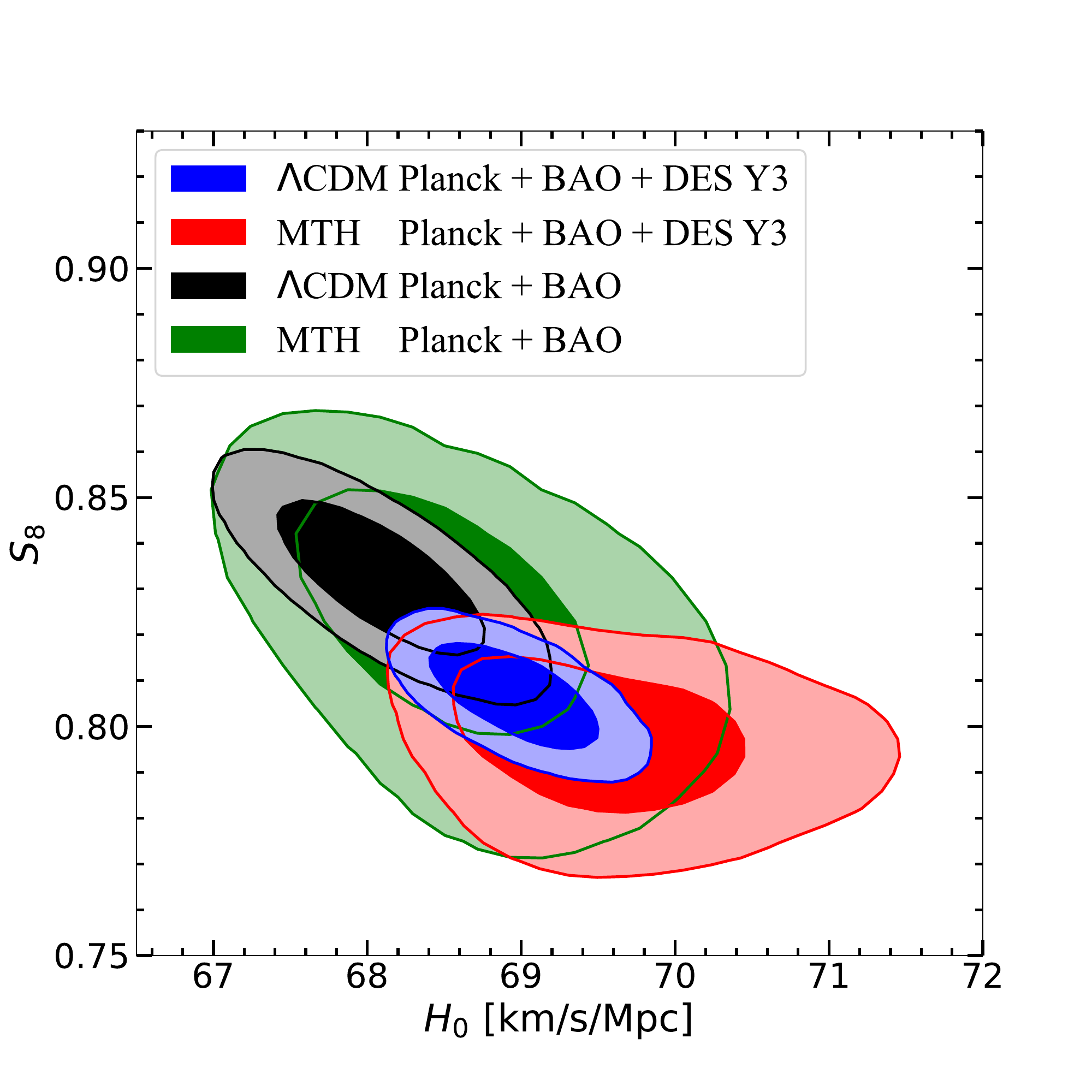}
    \includegraphics[scale = 0.35]{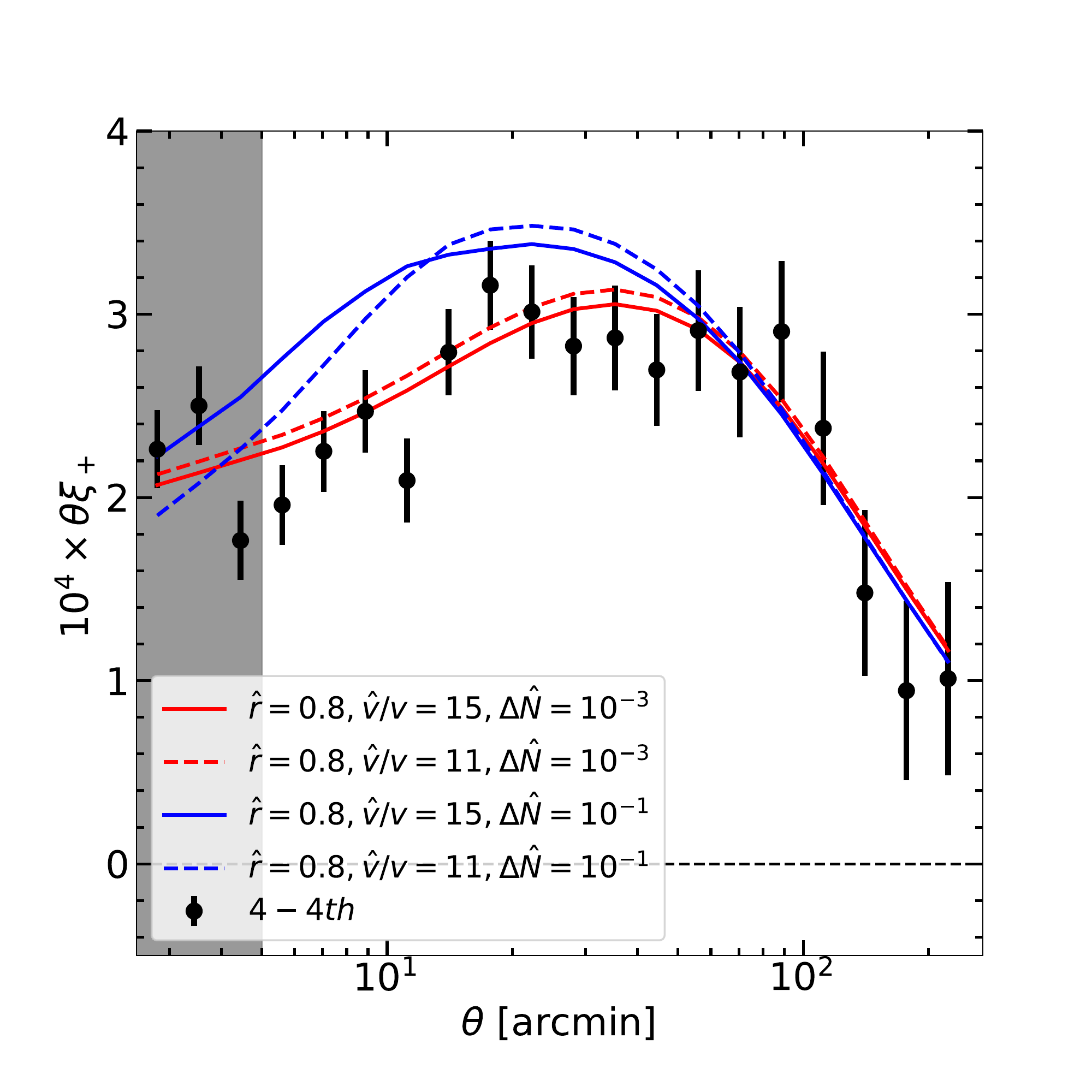}
    \includegraphics[scale = 0.35]{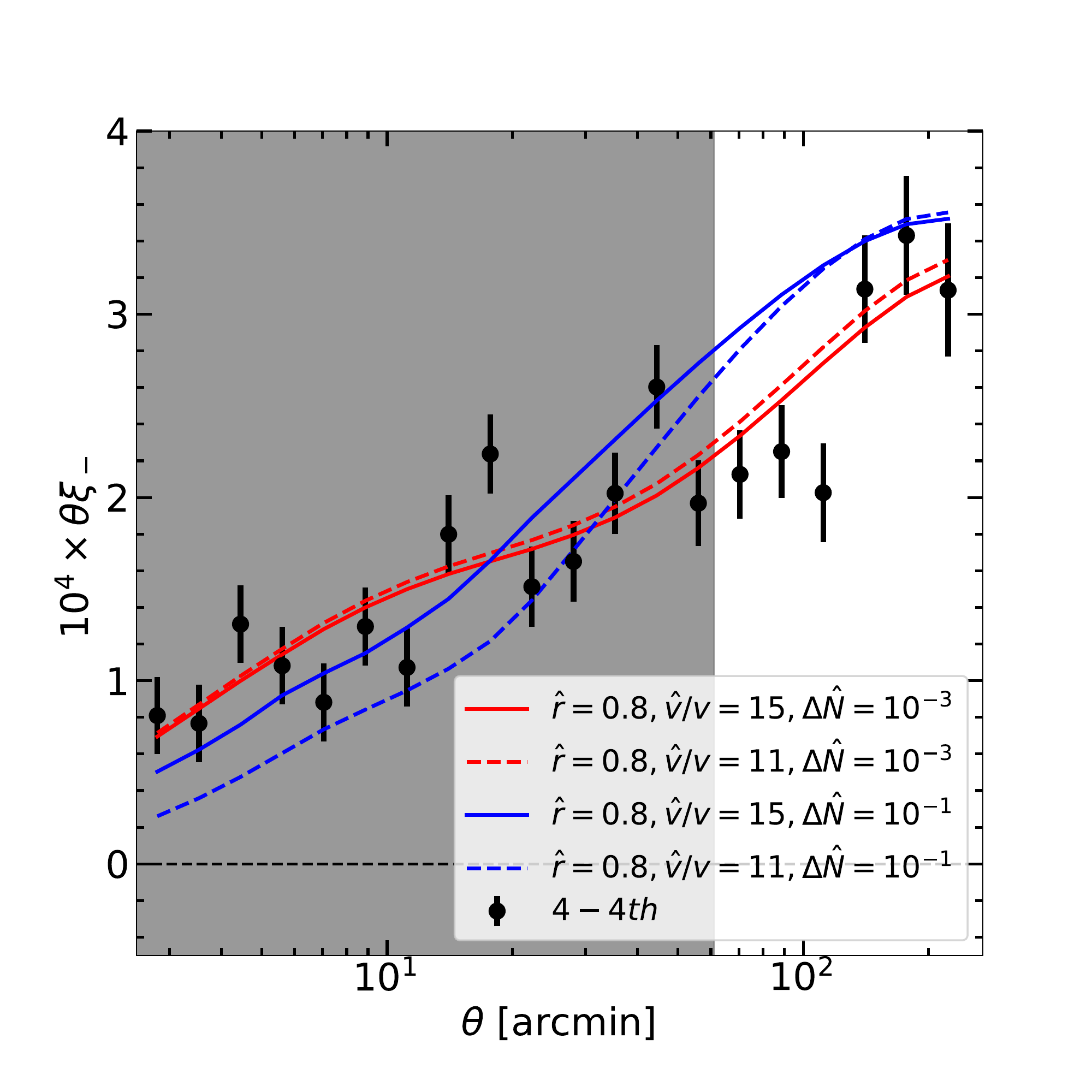}
    \caption{The two dimensional posterior distributions for MTH parameters $\hat{r}$ and $\Delta \hat{N}$ (upper left panel) 
    as well as cosmological parameters $H_0$ and $S_8$ (upper right panel). 
    The contours show the 68\% (inner contour) and 95\% (outer contour) credible regions.  
    The green contours are based on the likelihood of \texttt{Planck+BAO}, 
    while the likelihood for the red contours additionally includes the DES~Y3 cosmic shear data. 
    Here we use the \texttt{HMCode} for the non-linear correction as discussed in Sec.~\ref{sec:model} and \ref{sec:N_body}.
    The black (blue) contours are based on the likelihood of \texttt{Planck+BAO} (\texttt{Planck+BAO+DES Y3}) for the $\Lambda$CDM model.
    The two bottom graphs show the auto two-point correlation function of cosmic shear in 4-4th redshift bin for $\xi_+$ (bottom left panel) and $\xi_{-}$ (bottom right panel).
    The red lines denote the MTH parameters $\hat{r}=0.8$, $\Delta \hat{N}=0.001$ with $\hat{v}/v=15$ (solid line) and $\hat{v}/v=11$ (dashed line).
    The blue lines denote the MTH parameters $\hat{r}=0.8$, $\Delta \hat{N}=0.1$ with $\hat{v}/v=15$ (solid line) and $\hat{v}/v=11$ (dashed line).}
  
    
    \label{fig:r_dn_h0_s8_contour}
\end{figure}


\begin{table}
    \centering
    \begin{tabular}{|c|c|c|c|c|c|c|c|}
\hline \multirow[b]{2}{*}{ Parameters } & \multicolumn{2}{c|}{$\Lambda \mathrm{CDM}$} & \multicolumn{2}{c|}{ MTH } \\
\hline & best-fit & $\operatorname{mean} \pm \sigma$ & best-fit & $\operatorname{mean} \pm \sigma$ \\ 
\hline
$100 \Omega_b h^2$ & 2.262 & $2.256_{-0.0131}^{+0.0132}$ & 2.262 & $2.263_{-0.015}^{+0.014}$  \\
$\Omega_{\rm cdm} h^2$ & 0.1179 & $0.1177_{-0.00071}^{+0.00074}$ & 0.1192 & $0.1196_{-0.0023}^{+0.0014}$  \\
$100 \theta_s$ & 1.042 & $1.042_{-0.00028}^{+0.0003}$ & 1.042 & $1.042_{-0.00033}^{+0.00035}$  \\
$\ln \left(10^{10} A_s\right)$ & 3.023 & $3.034_{-0.014}^{+0.014}$ & 3.049 & $3.038_{-0.015}^{+0.015}$  \\
$n_s$ & 0.9713 & $0.9709_{-0.00349}^{+0.00353}$ & 0.9701 & $0.9729_{-0.0047}^{+0.0039}$  \\
$\tau_{\text {reio }}$ & 0.04851 & $0.05194_{-0.0071}^{+0.0071}$ & 0.05741 & $0.05231_{-0.0073}^{+0.0074}$  \\
\hline
$\hat{r}$ & -- & -- & 0.0375 & $0.1812_{-0.1802}^{+0.015}$   \\
$\hat{v} / v$ & -- & -- & 2.3625 & $7.707_{-5.707}^{+2.051}$  \\
$\Delta \hat N$ & -- & -- & 0.0435 & $0.1064_{-0.1054}^{+0.0319}$   \\
\hline
$\Omega_m$ & 0.2954 & $0.295_{-0.0043}^{+0.0041}$ & 0.2973 & $0.294_{-0.0051}^{+0.0049}$ \\ 
$H_0$ & 68.97 & $68.97_{-0.349}^{+0.344}$ & 69.06 & $69.55_{-0.76}^{+0.49}$  \\
$S_8$ & 0.8038 & $0.8067_{-0.0073}^{+0.0075}$ & 0.802 & $0.797_{-0.0093}^{+0.012}$  \\
\hline$-2\ln\mathcal{L}$ & \multicolumn{2}{c|}{3032.4} & \multicolumn{2}{c|}{3031.74} \\
\hline \texttt{Planck + BAO} & \multicolumn{2}{c|}{2791.28} & \multicolumn{2}{c|}{2789.06}  \\
\hline \texttt{DES Y3} & \multicolumn{2}{c|}{241.12} & \multicolumn{2}{c|}{242.68} \\
\hline
\end{tabular}
    \caption{The mean and best-fit values for the $\Lambda$CDM and MTH models obtained including the \texttt{Planck+BAO+DES~Y3} likelihoods.}
    \label{tab:bestfit_des}
\end{table}

We start with a comparison between two scans; 
one uses the likelihood only containing the Planck and BAO data (\texttt{Planck+BAO}), 
and the other one further includes the DES Y3 cosmic shear likelihood (\texttt{Planck+BAO+DES~Y3}). 
In the top left panel of Fig.~\ref{fig:r_dn_h0_s8_contour}, we present the 68\% (inner contour) and 95\% (outer contour) credible regions 
projected on the ($\hat{r}$, $\Delta\hat N$) plane. 
The likelihoods of green contours are \texttt{Planck+BAO}, while 
the red contours are based on the \texttt{Planck+BAO+DES Y3} likelihood.  

We can clearly see two interesting features in the plot by comparing the allowed region between 
\texttt{Planck+BAO} and \texttt{Planck+BAO+DES Y3}. 
First, for $\Delta\hat N>0.06$,  
$\hat{r}$ has an upper limit $\hat r\lesssim0.3$ for the $95\%$ credible region from the inclusion of the DES likelihood. The bound gets much stronger than the result in~\cite{Bansal:2021dfh} (Fig.~21) with only the \texttt{Planck+BAO} data.  The upper bound on $\hat r$ owes to the effect that 
the matter power spectrum predicted by the MTH model is suppressed due to the twin BAO  (Fig.~\ref{Fig:p_ratio}). 
The suppression increases with $\hat r$ for a given $(\Delta\hat N,\hat{v}/v)$, and if $\hat r$ becomes too large, the model can no longer fit the DES data well. Our prior has an upper bound $\hat{v}/v\leq 15$, which corresponds to a better than $\approx 1\%$ tuning of the model~\cite{Craig:2015pha}. For $\Delta\hat N>0.06$, even if $\hat{v}/v\approx15$, the twin recombination still occurs late enough so that the DES data is sensitive to the power spectrum suppression. Conversely, a smaller value of $\Delta\hat N$ corresponds to colder MTH particles and an earlier twin recombination process, resulting in the twin baryon behaving like CDM at an earlier time. At the bottom of Fig.\ref{fig:r_dn_h0_s8_contour}, it is shown that we can improve the fit of the cosmic shear data by decreasing $\Delta \hat{N}$ from 0.1 (blue line) to 0.001 (red line) for large $\hat{r}=0.8$. However, increasing the value of $\hat{v}/v$ does not significantly improve the fit, particularly within our prior range of $\hat{v}/v \leqslant 15$. Therefore, to relax the constraint on $\hat r$, a colder mirror sector is needed, which is reflected by a smaller value of $\Delta\hat N$ (as is also illustrated in Fig.~\ref{reweight_corner}).  
Another interesting feature in the plot is that when $\hat{r}<0.2$, the upper bounds on $\Delta\hat N$ from the red curves extend to higher values than the green curves. 
The weaker $\Delta\hat N$ bound indicates that the DES~Y3 data prefer a slightly warmer mirror sector that can result in more significant suppression of the matter power spectrum. 
Moreover, as discussed in Ref.~\cite{Bansal:2021dfh}, the presence of twin photons as scattering radiation (before the twin recombination) also weakens the $\Delta N_{\rm eff}$ constraints in the MTH model compared to the free-streaming radiation\footnote{See also~\cite{Baumann:2015rya,Brust:2017nmv, Blinov:2020hmc,Ghosh:2021axu,Brinckmann:2022ajr} for more studies of interacting DR models.}.

In the upper right panel of Fig.~\ref{fig:r_dn_h0_s8_contour}, we present the preferred region of $(H_0,S_8)$ from fitting the MTH (green and red) and $\Lambda$CDM (black and blue) models. 
Compared to the fit without the DES~Y3 data, both the $\Lambda$CDM and MTH contours move to smaller $S_8$ regions, indicating that the DES~Y3 data prefers a smaller matter power spectrum in the scale $0.1\lesssim k\lesssim 10\,h$Mpc$^{-1}$ than the model predictions from fitting the CMB and BAO data. In the MTH model, the mirror radiation and the twin acoustic oscillation process can enhance $H_0$ while keeping a small $S_8$, providing a chance to relax both tensions. Notice that the MTH model has a limit to the $\Lambda$CDM model either with $\hat r\to 0$ or $\Delta\hat N\to 0$. This means that the MTH contours contain the best-fit points of the $\Lambda$CDM model, and the larger MTH contours do mean that the model can accommodate a larger $H_0$ and a smaller $S_8$ compared to the $\Lambda$CDM. The red contour in the plot has $H_0=69.56_{-0.76}^{+0.49}$ km s$^{-1}$Mpc$^{-1}$, and the Gaussian Tension to the SH0ES result is $2.8\sigma$ compared to the $\Lambda$CDM result $H_0=69.0_{-0.35}^{+0.34}$ km s$^{-1}$Mpc$^{-1}$ and $3.7\sigma$. Although the Gaussian Tension is only a rough measure of disagreement between measurements, the numbers indeed show an improvement in relaxing the $H_0$ tension. 

\subsection{Likelihoods including SH0ES}
\label{sec:SH0ES}

\begin{figure}[t!]
    \centering
    \includegraphics[scale = 0.4]{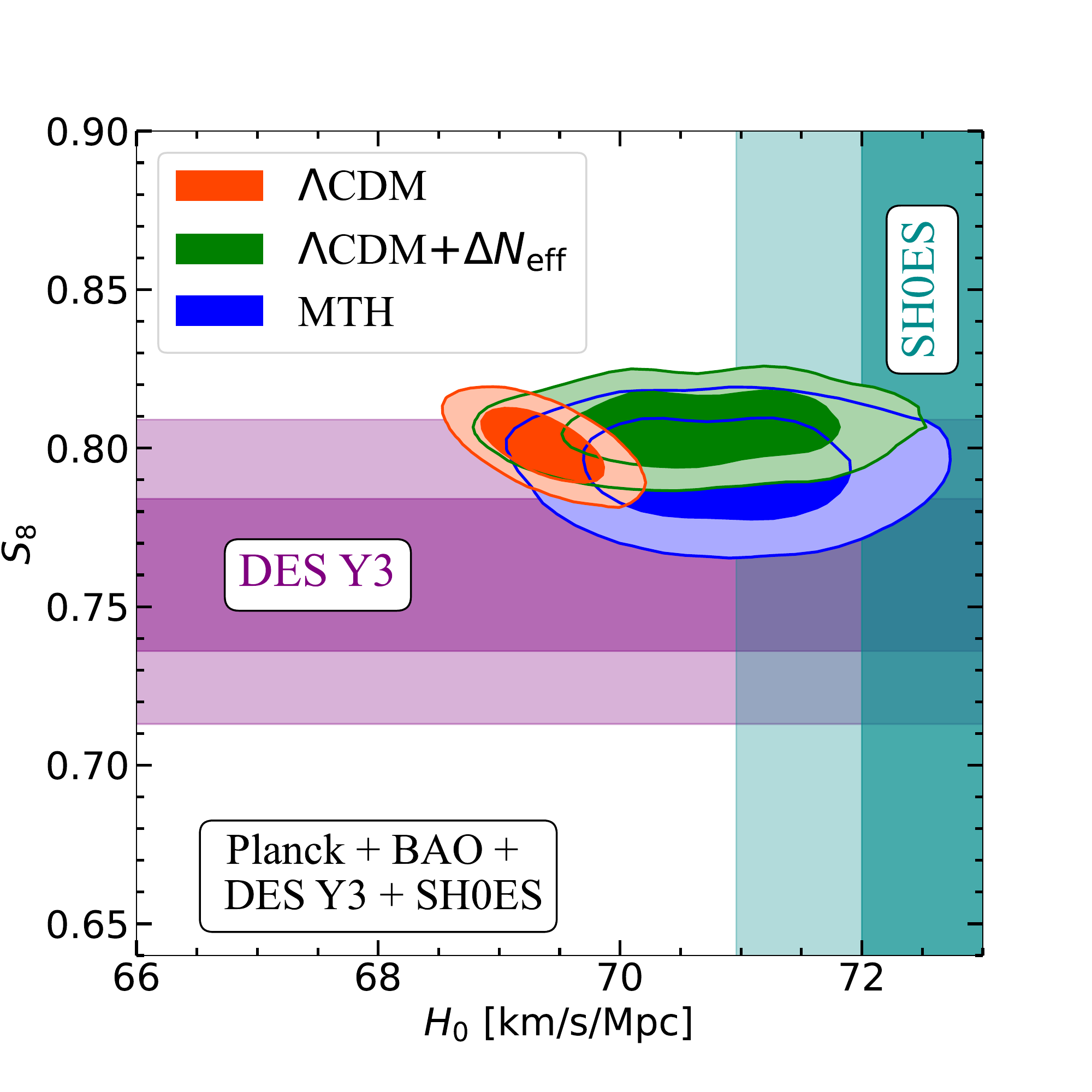}
    \caption{Marginalized 2D posterior distribution in the ($H_0$ , $S_8$) plane with \texttt{Planck+BAO+DES Y3+SH0ES}. The orange contour corresponds to the $\Lambda \rm{CDM}$ model, the green contour corresponds to the $\Lambda \rm{CDM} + \Delta N_{\rm eff}$ model and the blue contour corresponds to the MTH model.
    The dark cyan band corresponds to the $1\sigma$~(inner) and $2\sigma$~(outer) region of the SH0ES result~\cite{Riess:2021jrx}, and the purple band corresponds to the $S_8$ obtained by a blind analysis in the context of the $\Lambda$CDM model~\cite{DES:2021bvc}.
    }
    \label{fig:h0_s8_planck_DES_SH0ES}
\end{figure}
\begin{table}
    \centering
    \begin{tabular}{|c|c|c|c|c|c|c|c|}
\hline \multirow[b]{2}{*}{ Parameters } & \multicolumn{2}{c|}{$\Lambda \mathrm{CDM}$} & \multicolumn{2}{c|}{$\Lambda \mathrm{CDM} + \Delta N_{\rm eff}$} & \multicolumn{2}{c|}{ MTH } \\
\hline & best-fit & $\operatorname{mean} \pm \sigma$ & best-fit & $\operatorname{mean} \pm \sigma$ & best-fit & $\operatorname{mean} \pm \sigma$ \\ 
\hline
$100 \Omega_b h^2$ & 2.270 & $2.268_{-0.0148}^{+0.0099}$ & 2.275 & $2.282_{-0.0175}^{+0.0115}$  & 2.273 & $2.283_{-0.017}^{+0.012}$  \\
$\Omega_{\rm cdm} h^2$ & 0.1166 & $0.1167_{-0.0005}^{+0.0009}$ & 0.1204 & $0.1211_{-0.0026}^{+0.017}$  & 0.1210 & $0.1224_{-0.0028}^{+0.0021}$  \\
$100 \theta_s$ & 1.042 & $1.042_{-0.0003}^{+0.0002}$ & 1.042 & $1.042_{-0.00032}^{+0.00046}$  & 1.042 & $1.042_{-0.00035}^{+0.00042}$  \\
$\ln \left(10^{10} A_s\right)$ & 3.048 & $3.044_{-0.016}^{+0.012}$ & 3.052 & $3.05_{-0.016}^{+0.012}$  & 3.054 & $3.051_{-0.016}^{+0.013}$  \\
$n_s$ & 0.9735 & $0.9736_{-0.0041}^{+0.0028}$ & 0.9803 & $0.9811_{-0.0062}^{+0.0038}$  & 0.9740 & $0.9795_{-0.0054}^{+0.0036}$  \\
$\tau_{\text {reio }}$ & 0.0595 & $0.0572_{-0.0082}^{+0.0060}$ & 0.05784 & $0.0555_{-0.0074}^{+0.0067}$  & 0.05866 & $0.05566_{-0.0074}^{+0.0068}$  \\
\hline
$\hat{r}$ & -- & -- & -- & -- & 0.1144 & $0.1128_{-0.076}^{+0.091}$   \\
$\hat{v} / v$ & -- & -- & -- & -- & 9.62 & $8.98_{-4.04}^{+3.76}$  \\
$\Delta\hat N$ & -- & -- & 0.2071 & $0.2588_{-0.1545}^{+0.093}$ & 0.1979 & $0.3098_{-0.1520}^{+0.0950}$ \\
\hline
$\Omega_m$ & 0.2908 & $0.2901_{-0.0040}^{+0.0039}$ & 0.288 & $0.2873_{-0.0041}^{+0.0041}$ & 0.2915 & $0.2895_{-0.005}^{+0.0046}$ \\ 
$H_0$ & 69.51 & $69.48_{-0.43}^{+0.21}$ & 70.5 & $70.65_{-0.732}^{+0.775}$ & 70.22 & $71.04_{-0.92}^{+0.52}$  \\
$S_8$ & 0.8159 & $0.8145_{-0.0055}^{+0.0052}$ & 0.8086 & $0.8062_{-0.0078}^{+0.0078}$ & 0.8034 & $0.794_{-0.0099}^{+0.0117}$  \\
\hline$-2\ln\mathcal{L}$ & \multicolumn{2}{c|}{3046.49} & \multicolumn{2}{c|}{3042.76} & \multicolumn{2}{c|}{3042.21} \\
\hline \texttt{Planck+BAO} & \multicolumn{2}{c|}{2794.87} & \multicolumn{2}{c|}{2794.27} & \multicolumn{2}{c|}{2792.9}  \\
\hline \texttt{DES Y3} & \multicolumn{2}{c|}{240.09} & \multicolumn{2}{c|}{242.52} & \multicolumn{2}{c|}{241.81} \\
\hline \texttt{SH0ES} & \multicolumn{2}{c|}{11.53} & \multicolumn{2}{c|}{5.966} & \multicolumn{2}{c|}{7.355} \\
\hline
\end{tabular}
    \caption{The mean and best-fit values for the $\Lambda$CDM, $\Lambda$CDM+$\Delta N_{\rm eff}$, and MTH models obtained including the \texttt{Planck+BAO+DES Y3+SH0ES} likelihoods.}
    \label{tab:bestfit_shoes}
\end{table}
In order to test whether the MTH model could alleviate $H_0$ tension and $S_8$ tensions compared to the $\Lambda$CDM, we have included the \texttt{SH0ES} data in our fitting.
In Fig.~\ref{fig:h0_s8_planck_DES_SH0ES}, we show the ($H_0, S_8$) results based on three different models: $\Lambda$CDM, $\Lambda \rm{CDM}+\Delta N_{\rm{eff}}$ and MTH. For comparison, we show the $1$ and $2\sigma$ preferred region of the $S_8$ obtained by a blind analysis in the context of the $\Lambda$CDM model ~\cite{DES:2021bvc} (purple band). We also show the $1$ and $2\sigma$ preferred regions from the SH0ES measurement~\cite{Riess:2021jrx} (dark cyan band). The $\Lambda$CDM model is in significant tension to both measurements, especially the SH0ES. When using the $\Lambda$CDM+$\Delta N_{\rm{eff}}$ model to reduce the $H_0$ tension, the tension in $S_8$ gets a bit worse than the $\Lambda$CDM result. On the other hand, the MTH contour can extend to a larger $H_0$ while having a smaller $S_8$ value. This indicates that the MTH model has the potential to alleviate both $H_0$ and $S_8$ tensions, rather than worsening the $S_8$ tension as the  $\Lambda\rm{CDM}+\Delta N_{\rm{eff}}$ model. However, since the $S_8$ tension in the DES~Y3 data is not as significant as the SH0ES $H_0$ tension, the likelihood ratio $-2\ln(\mathcal{L}_{\rm MTH}/\mathcal{L}_{\Lambda\rm{CDM}})=-4.28$ is only slightly lower than  $-2\ln(\mathcal{L}_{\Lambda\rm{CDM}+\Delta N_{\rm{eff}}}/\mathcal{L}_{\Lambda\rm{CDM}})=-3.73$. Given that the MTH model has three more parameters than $\Lambda$CDM, the model does not perform better than the $\Lambda\rm{CDM}+\Delta N_{\rm{eff}}$ model in fitting the \texttt{Planck+BAO+DES~Y3+SH0ES} data.\footnote{A recent study~\cite{Rubira:2022xhb} investigated the interacting dark matter (IDM) model and found that including the measurement of $f\sigma_8$ from redshift space distortion (RSD) of BOSS DR12 does not allow the model to have both a small $S_8$ and large $\Delta N_{\rm eff}$ in the fit. See also a relevant study in~\cite{Raveri_2017}. Although the IDM model differs from the MTH model, which has the twin recombination process to alleviate the strong correlation between the $\Delta N_{eff}$ enhancement and the suppression of the power spectrum, both models comprise a sub-component of DM scattering with DR and demonstrate a similar dark acoustic oscillation process. We leave a closer comparison of the two models for future study.}

Suppose the future $S_8$ measurements converge to the mean value of the DES~Y3 result (center of the purple band in Fig.~\ref{fig:h0_s8_planck_DES_SH0ES}), but with the error drops by a factor of two. In that case, the $S_8$ will be close to the Planck~SZ~(2013) result, and the significance of the $S_8$ tension will be comparable to the current $H_0$ problem. Based on the scenario, we will show that the MTH model can provide a much better fit to the data if the $S_8$ tension worsens.

\subsection{Likelihoods including Planck~SZ (2013)}
\label{sec:SZ_SH0ES}

\begin{figure}[t!]
    \centering
    \includegraphics[scale = 0.4]{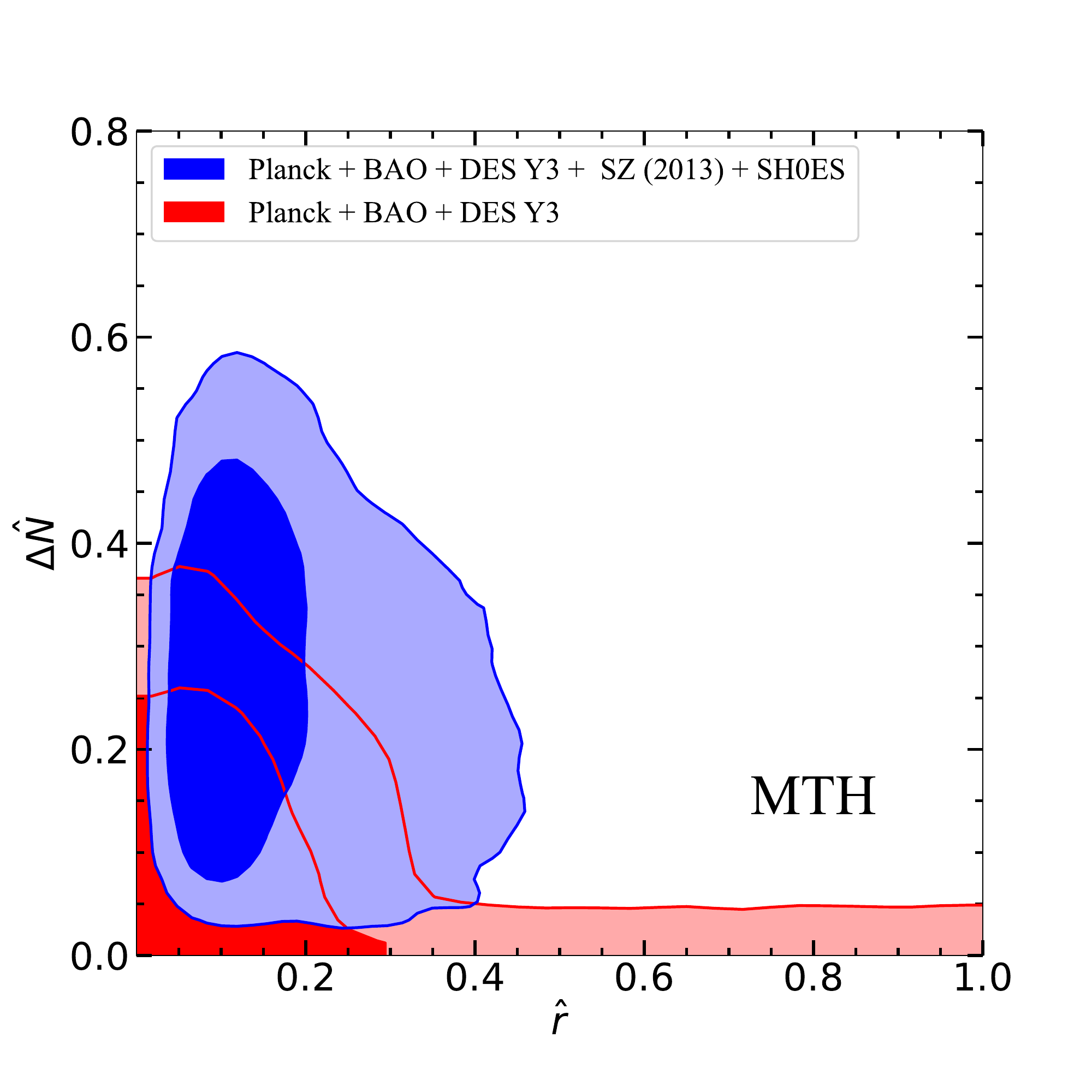}
    \caption{Marginalized posterior distribution similar to Fig.~\ref{fig:r_dn_h0_s8_contour} but  
    for a comparison between \texttt{Planck+BAO+DES~Y3} (red contours) and 
    \texttt{Planck+BAO+DES~Y3+SZ~(2013)+SH0ES} (blue contours). 
    }
    \label{fig:r_dn_sz_contour}
\end{figure}

\begin{figure}[t!]
    \centering
    \includegraphics[scale = 0.40]{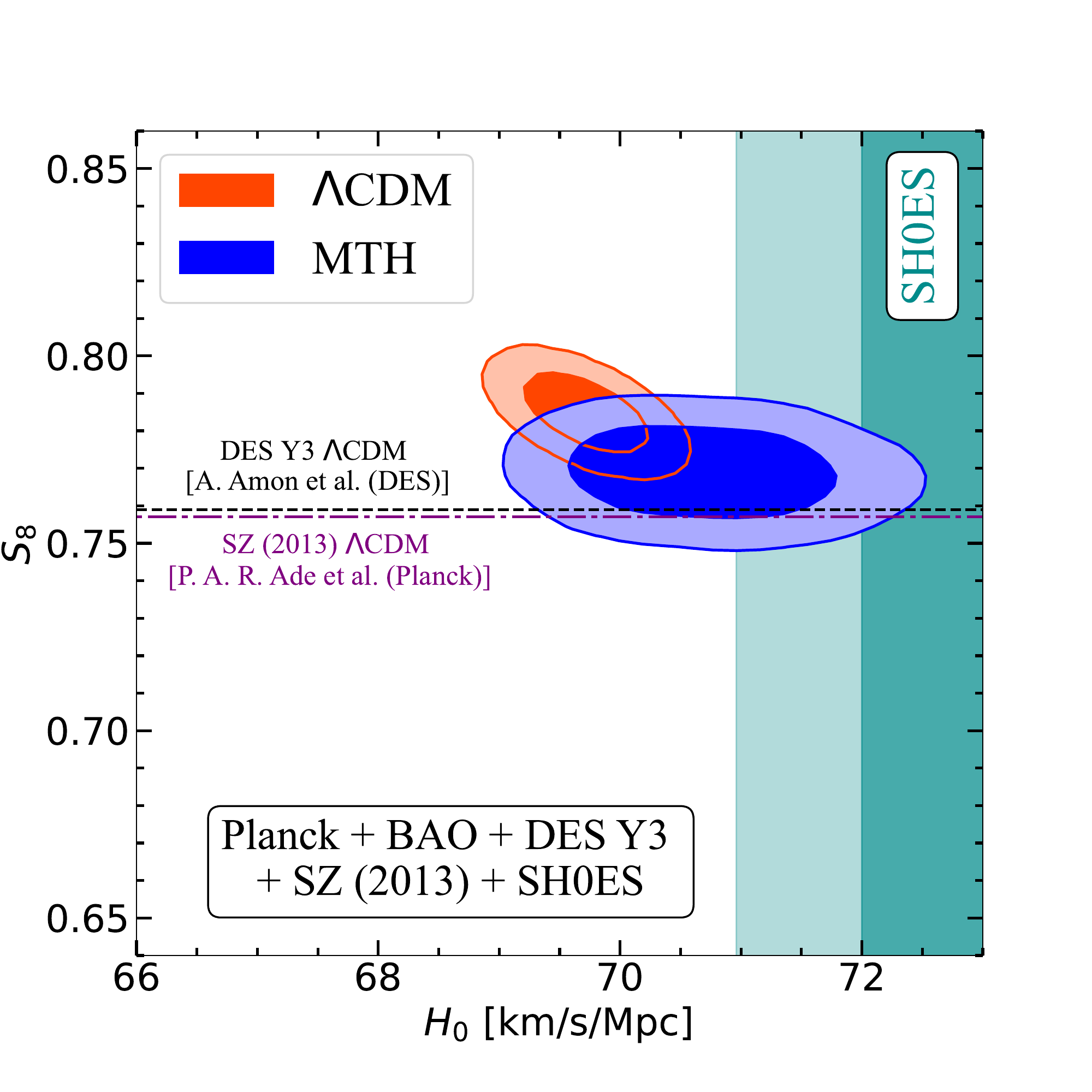}
    \caption{Marginalized posterior distribution of the scanning scenario 
    \texttt{Planck+BAO+DES Y3+SZ~(2013)+SH0ES} for $\Lambda$CDM (orange contour) 
    and MTH (blue contour). 
    The dark cyan band corresponds to the $95\%$ region of $H_0$ measured by SH0ES collaboration.
    Two dashed lines denote the central value of $S_8$ from DES Y3 and Planck~SZ (2013) measurement.
    }
    \label{fig:h0_s8_sz_LCDM}
\end{figure}

In Fig.~\ref{fig:r_dn_sz_contour}, we compare the credible regions of 
\texttt{Planck+BAO+DES~Y3} (red contours) and \texttt{Planck+ BAO+DES Y3+SZ (2013)+SH0ES} (blue contours) 
for the MTH. The worse $S_8$ tension in the Planck~SZ~(2013) data than in the DES~Y3 pushes the upper bound on $\hat r$ to a higher value. Moreover, both the $\hat{r}$ and $\Delta\hat N$ in this case have non-zero lower limits of the $95\%$ credible region, namely 
$0.02 \lesssim \hat{r} \lesssim 0.4$ and $0.03 \lesssim \Delta\hat N\lesssim 0.5$. Hence, the \texttt{SZ~(2013)+SH0ES} likelihood prefers a non-vanishing twin matter component with a high enough temperature to sustain the twin BAO process. On the other hand, the suppression of the matter power spectrum cannot be too significant to violate 
the cosmic shear constraints. 
Hence, the dataset favors a particular range of non-zero $\hat{r}$. 
Similar results have also been shown in Ref.~\cite{Bansal:2021dfh}. 
However, we find the upper limits for $\hat{r}$ as shown in Fig.~\ref{fig:r_dn_sz_contour} for the first time. In Fig.~\ref{fig:h0_s8_sz_LCDM}, we show a comparison between the $\Lambda$CDM (orange contour) and 
MTH (blue contour) based on the likelihood scenario \texttt{Planck+BAO+DES Y3+SZ~(2013)+SH0ES}. Although the MTH model has a continuous limit ($\hat r\to 0$ and $\Delta\hat N\to 0$) to reproduce the fit of $\Lambda$CDM model, the dataset pushes the MTH contour to a much lower $S_8$ and larger $H_0$ region. Although $\Lambda$CDM+$\Delta N_{\rm eff}$ model can also accommodate a larger $H_0$, it does a worse job in resolving the $S_8$ tension. Compared to the +$\Delta N_{\rm eff}$ model, the MTH can further reach the region that resolves both tensions.

\begin{table}
    \centering
    \begin{tabular}{|c|c|c|c|c|c|c|c|}
\hline \multirow[b]{2}{*}{ Param } & \multicolumn{2}{c|}{$\Lambda \mathrm{CDM}$} & \multicolumn{2}{c|}{ MTH } \\
\hline & best-fit & $\operatorname{mean} \pm \sigma$ & best-fit & $\operatorname{mean} \pm \sigma$ \\ \hline
$100 \Omega_b h^2$ & 2.272 & $2.273_{-0.015}^{+0.011}$ & 2.286 & $2.285_{-0.018}^{+0.012}$  \\
$\Omega_{\rm cdm} h^2$ & 0.1164 & $0.1159_{-0.00052}^{+0.00089}$ & 0.1205 & $0.1226_{-0.0028}^{+0.0020}$  \\
$100 \theta_s$ & 1.042 & $1.042_{-0.00031}^{+0.00025}$ & 1.042 & $1.042_{-0.00034}^{+0.00040}$  \\
$\ln \left(10^{10} A_s\right)$ & 3.03 & $3.026_{-0.016}^{+0.011}$ & 3.034 & $3.046_{-0.016}^{+0.013}$  \\
$n_s$ & 0.9749 & $0.9744_{-0.0041}^{+0.0029}$ & 0.9747 & $0.9772_{-0.0051}^{+0.0034}$  \\
$\tau_{\text {reio }}$ & 0.05286 & $0.05_{-0.0082}^{+0.0059}$ & 0.05058 & $0.05419_{-0.0076}^{+0.0068}$  \\
\hline
$\hat{r}$ & -- & -- & 0.06745 & $0.1537_{-0.074}^{+0.087}$   \\
$\hat{v} / v$ & -- & -- & 2.725 & $5.968_{-2.513}^{+2.329}$  \\
$\Delta\hat N$ & -- & -- & 0.1676 & $0.3002_{-0.1449}^{+0.0898}$ \\
\hline
$\Omega_m$ & 0.2867 & $0.2856_{-0.0039}^{+0.0040}$ & 0.2908 & $0.29_{-0.0050}^{+0.0051}$ \\ 
$H_0$ & 69.65 & $69.72_{-0.34}^{+0.33}$ & 70.23 & $70.71_{-0.72}^{+0.70}$  \\
$S_8$ & 0.7896 & $0.7847_{-0.0065}^{+0.0065}$ & 0.7631 & $0.769_{-0.0082}^{+0.0082}$  \\
\hline $-2\ln\mathcal{L}$ & \multicolumn{2}{c|}{3063.72} & \multicolumn{2}{c|}{3048.68} \\
\hline \texttt{Planck + BAO} & \multicolumn{2}{c|}{2798.51} & \multicolumn{2}{c|}{2801.012}  \\
\hline \texttt{DES Y3} & \multicolumn{2}{c|}{238.2} & \multicolumn{2}{c|}{239.3} \\
\hline \texttt{SZ (2013)} & \multicolumn{2}{c|}{16.39} & \multicolumn{2}{c|}{1.068}  \\
\hline \texttt{SH0ES} & \multicolumn{2}{c|}{10.625} & \multicolumn{2}{c|}{7.3} \\
\hline
\end{tabular}
    \caption{The mean and best-fit values for the $\Lambda$CDM and MTH models obtained including the \texttt{Planck+BAO+DES~Y3+SH0ES+SZ~(2013)} likelihoods.}
    \label{tab:bestfit}
\end{table}

To further show that the MTH provides a much better fit to this dataset, we show the best-fit parameters and the associated likelihood $-2\mathcal{L}$ of the two models in Table.~\ref{tab:bestfit}. 
The likelihood ratio  $-2\ln(\mathcal{L}_{\rm MTH}/\mathcal{L}_{\Lambda\rm{CDM}})=-15.04$ implies that the MTH indeed fits the likelihoods better even if the model has three more parameters. Adding three extra degrees of freedom is also statistically unimportant because the total number of degrees of freedom is much larger. The likelihood improvement mainly comes from the better fit of the SH0ES and Planck~SZ data. Compared with the MTH fit in Table.~\ref{tab:bestfit_des} without \texttt{SH0ES+SZ~(2013)}, we find that most of the best-fitted $\Lambda$CDM parameters change mildly. 
However, the MTH model parameters vary substantially between the two cases, which implies that the MTH sector is essential in alleviating two tensions.

\subsection{Forecast of DM detection from the CSST sensitivity} 
\label{sec:csst}
If the tension in the $H_0$ value persists and the $S_8$ parameter converges to the Planck~SZ (2013) result, as discussed in the previous section, the MTH model predicts non-zero values for the twin baryon and twin radiation energy densities when fit to cosmological data. This raises the interesting question of how precisely we can measure the abundance of twin particles with future experiments and surveys, assuming that the MTH model accurately describes the universe.

To address this question, we use the expected sensitivity of the upcoming CSST experiment to project the uncertainties in measuring the MTH parameters. Taking the same analysis procedure performed for DES Y3 in Sec.~\ref{sec:shear}, 
we mock the future CSST likelihood by taking the central values of the data ($D$) in Eq.~\eqref{eq:cslike} to be the shear-shear correlation function calculated based on the MTH parameters
\{$\Omega_b h^2 = 0.02281$, $\Omega_{\rm cdm} h^2=0.1237$, $\theta_s=1.042$, $\rm{ln}(10^{10}A_s)=3.037$, $n_s=0.9774$, $\tau_{\rm reio}=0.05094$, $\hat{r} = 0.1047$, $\hat{v}/v = 4.791$, $\Delta\hat N = 0.3071$\}. The numbers are within the $1\sigma$ region of the fit shown in Table.~\ref{tab:bestfit}. When calculating Eq.~\eqref{eq:cslike}, we adopt a relatively conservative per-component shape dispersion of the covariance matrix, $\sigma_e=0.3$~\cite{Gong:2019yxt}. Using the redshift distributions in Ref.~\cite{Lin:2022aro}, we perform an MCMC scan by encoding the mock CSST data in \texttt{Planck+BAO+CSST+SZ~(2013)+SH0ES} likelihood.

\begin{figure}[t!]
    \centering
    \includegraphics[scale = 0.37]{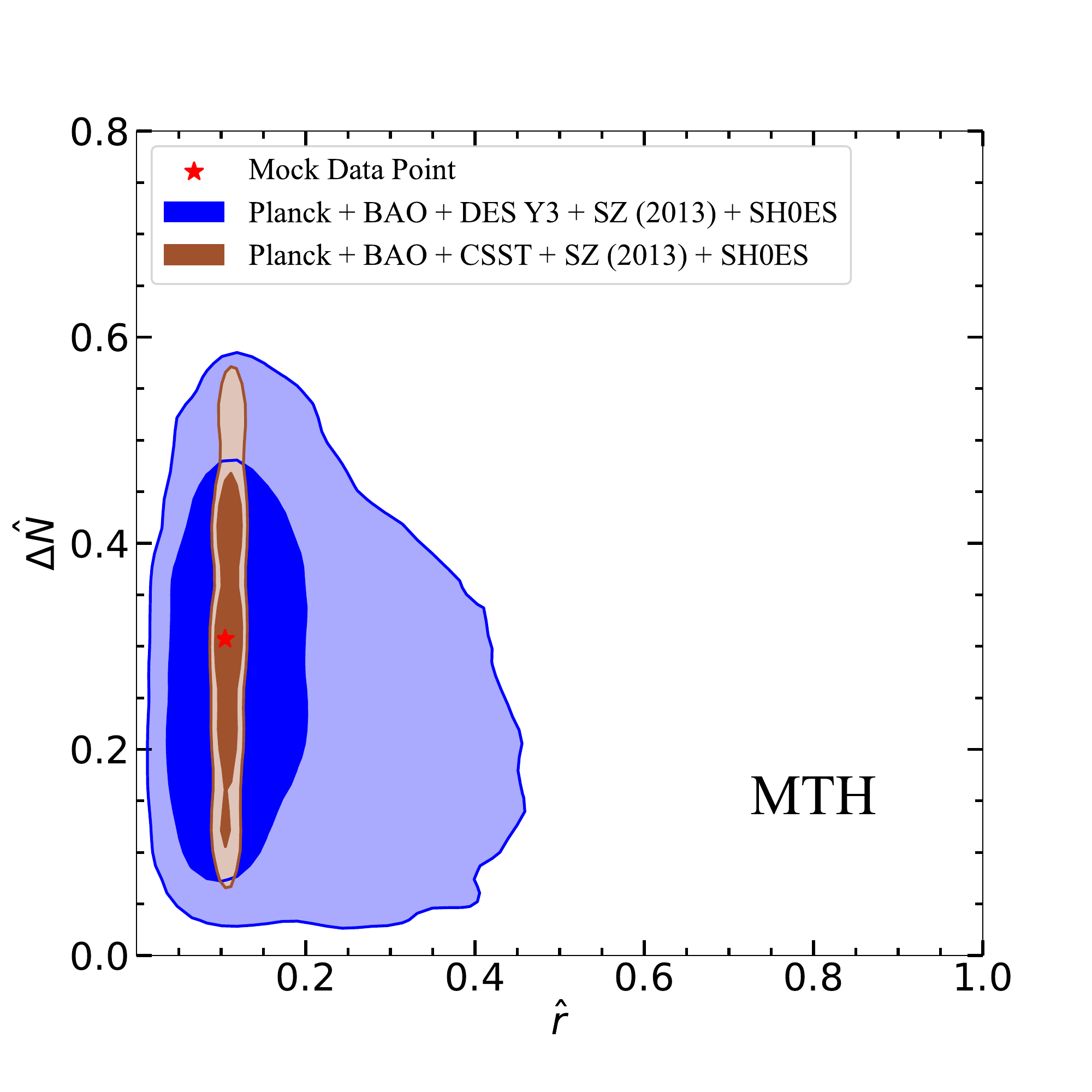}
    \includegraphics[scale = 0.37]{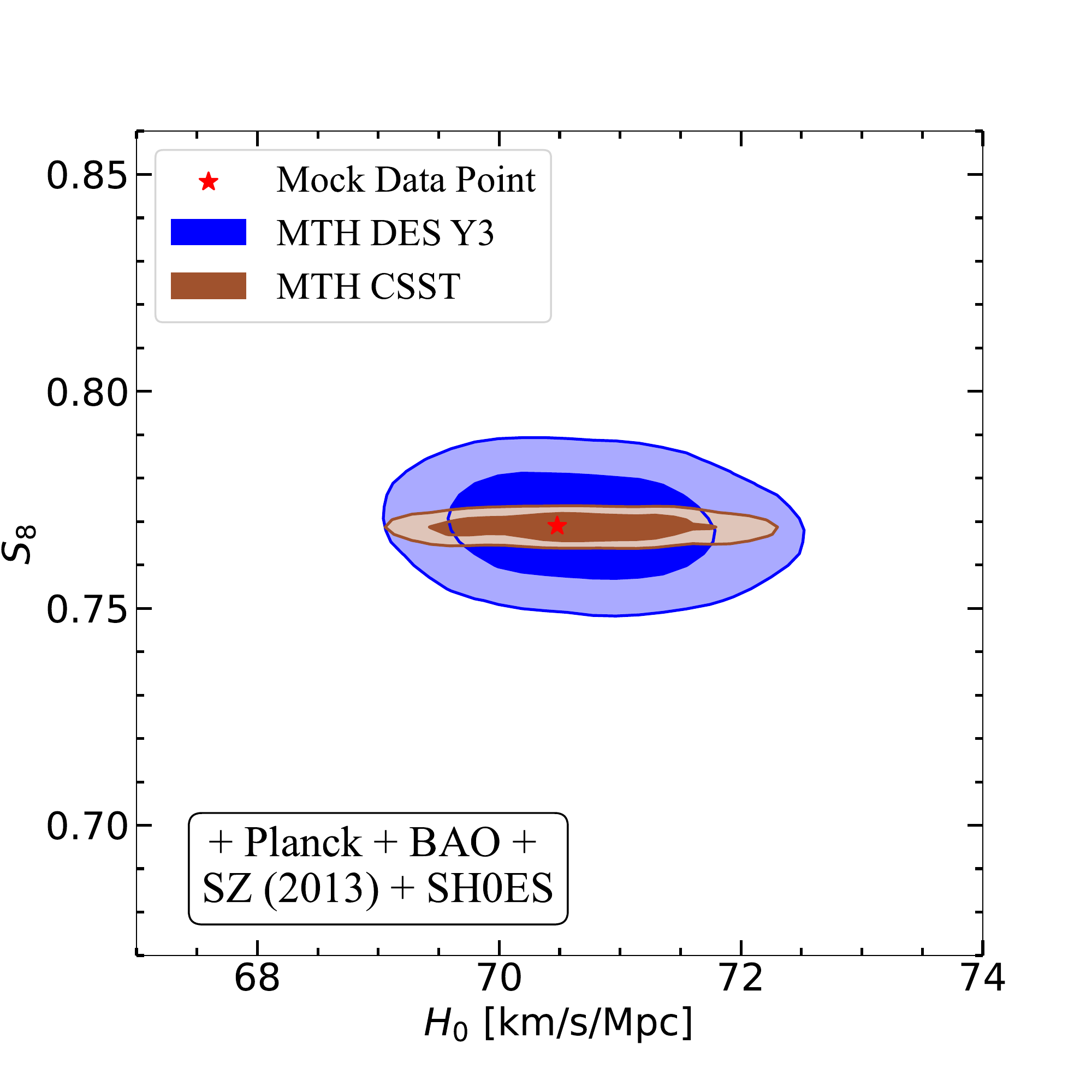}
    \caption{Marginalized 2D posterior distribution in the ($\hat{r}$ , $\Delta\hat N$) plane (left panel) and 
    the the ($H_0$ , $S_8$) plane (right panel), assuming a mock data point (red star):\{$\Omega_b h^2 = 0.02281$, $\Omega_{\rm cdm} h^2=0.1237$, $\theta_s=1.042$, $\rm{ln}(10^{10}A_s)=3.037$, $n_s=0.9774$, $\tau_{\rm reio}=0.05094$, $\hat{r} = 0.1047$, $\hat{v}/v = 4.791$, $\Delta\hat N = 0.3071$\}.
    The brown contours are the MTH result of \texttt{Planck+BAO+CSST+SZ (2013)+SH0ES} in comparison to  
    the \texttt{Planck+BAO+DES Y3+SZ (2013)+SH0ES} for the MTH (blue contours).
    Two dashed lines denote the central value of $S_8$ from DES Y3 and Planck~SZ (2013) measurement.
    }
    \label{fig:r_dn_h0_s8_sz_csst}
\end{figure}

The results of the MTH model between two scenarios before and after updating the future CSST cosmic shear likelihood  
are presented in the left panel of Fig.~\ref{fig:r_dn_h0_s8_sz_csst}. 
Attributing to a larger effective area and more galaxies to be observed in the CSST, 
the uncertainties of $\xi_{\pm}$ are significantly reduced at most a factor $\sim 10$. 
Hence, the most anticipated feature is the $95\%$ credible regions on the ($\hat{r}$, $\Delta\hat N$) plane (left panel) and ($H_0$ , $S_8$) plane (right panel) that significantly shrunk to the region near the central value, providing a $10\%$ ($1\%$) level precision measurement of $0.09<\hat{r}<0.12$ ($0.764<S_8<0.773$).

\section{Non-linear corrections:  \texttt{HMCode}, \texttt{Halofit} vs. $N$-body simulation}
\label{sec:N_body}
In this section, we justify the use of \texttt{HMCode} in our MCMC analysis by comparing the non-linear power spectra of two benchmark MTH models computed  
using \texttt{HMCode} and the gravity only $N$-body simulation.

\begin{figure}[h]
    \centering
    \includegraphics[scale = 0.37]{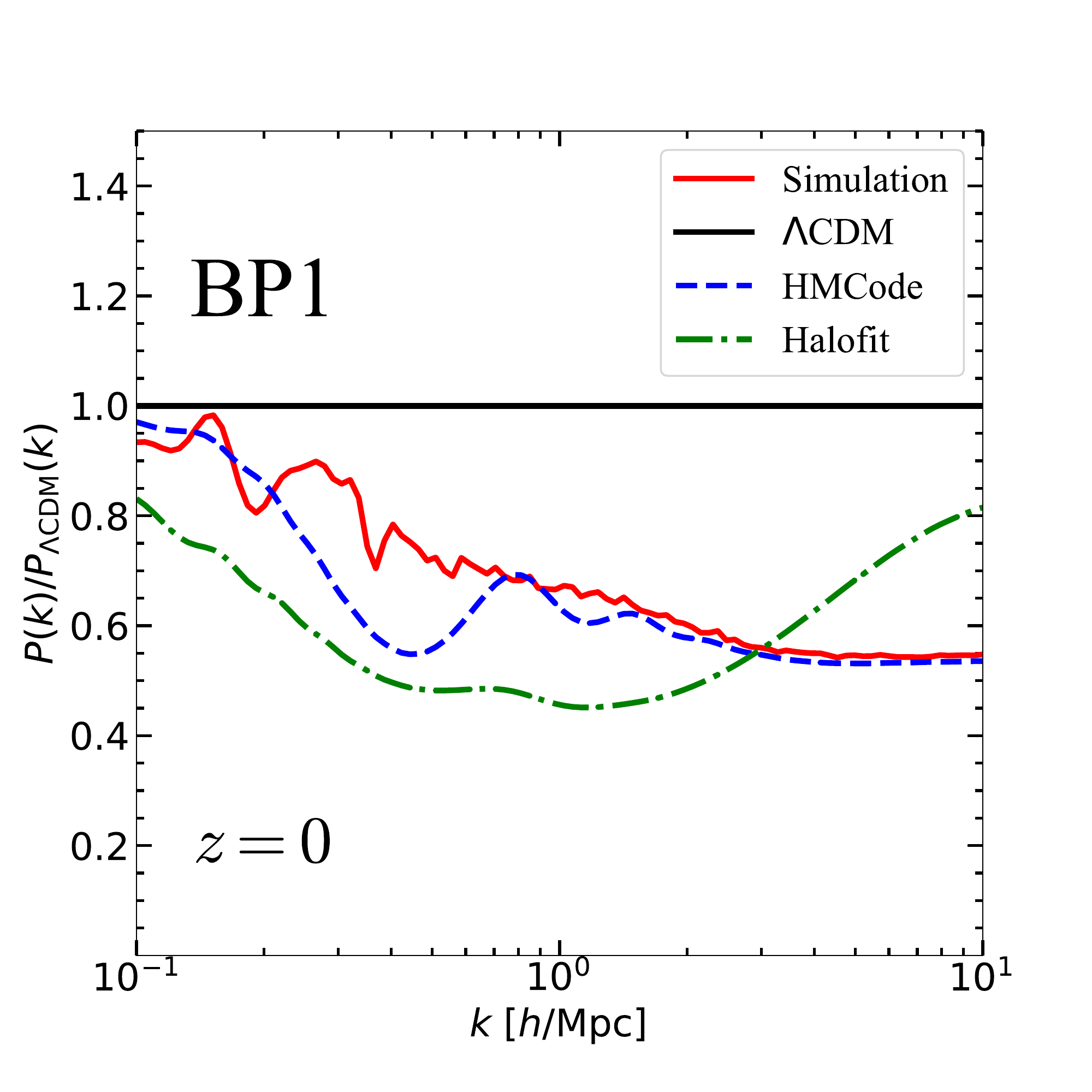}
    \includegraphics[scale = 0.37]{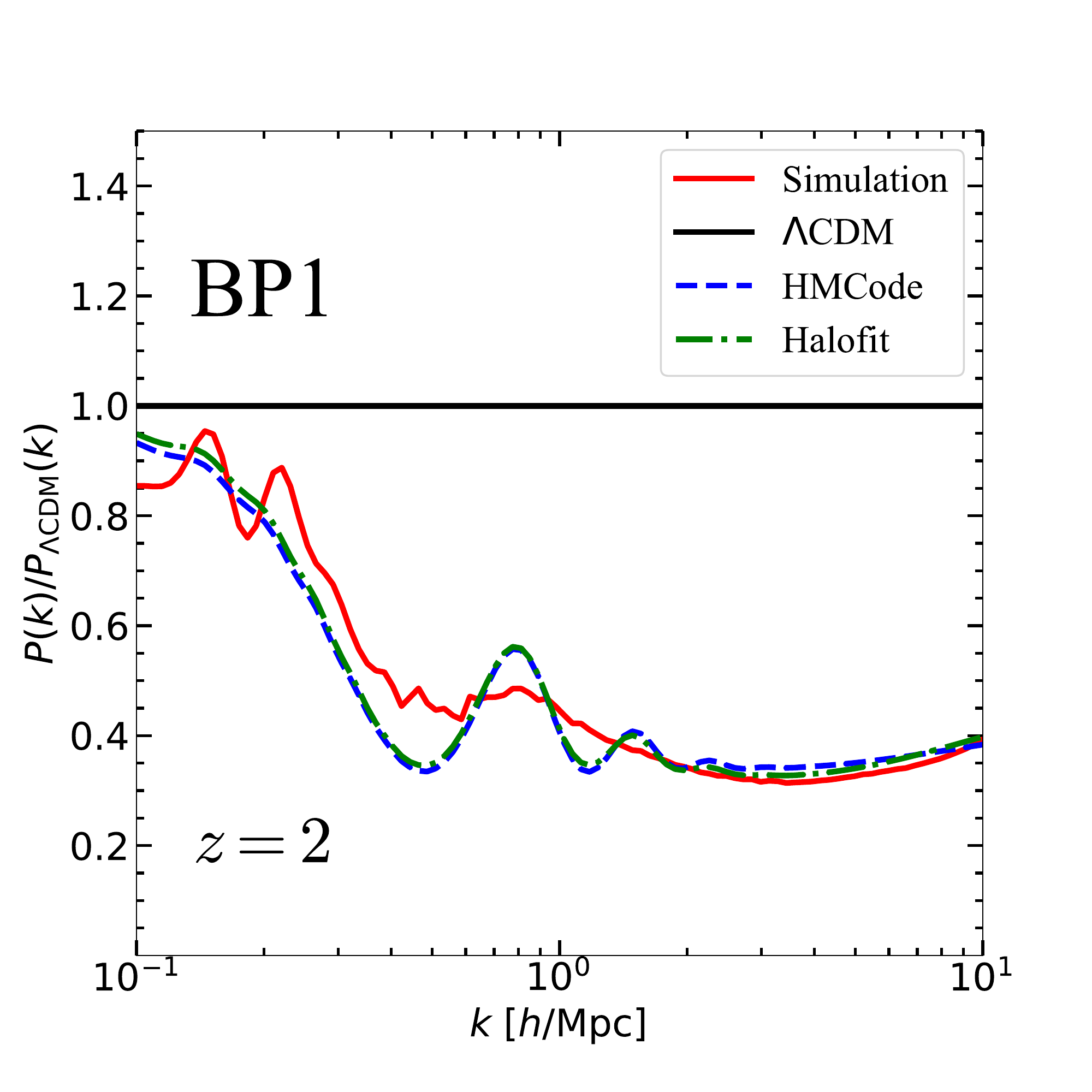}
    \includegraphics[scale = 0.37]{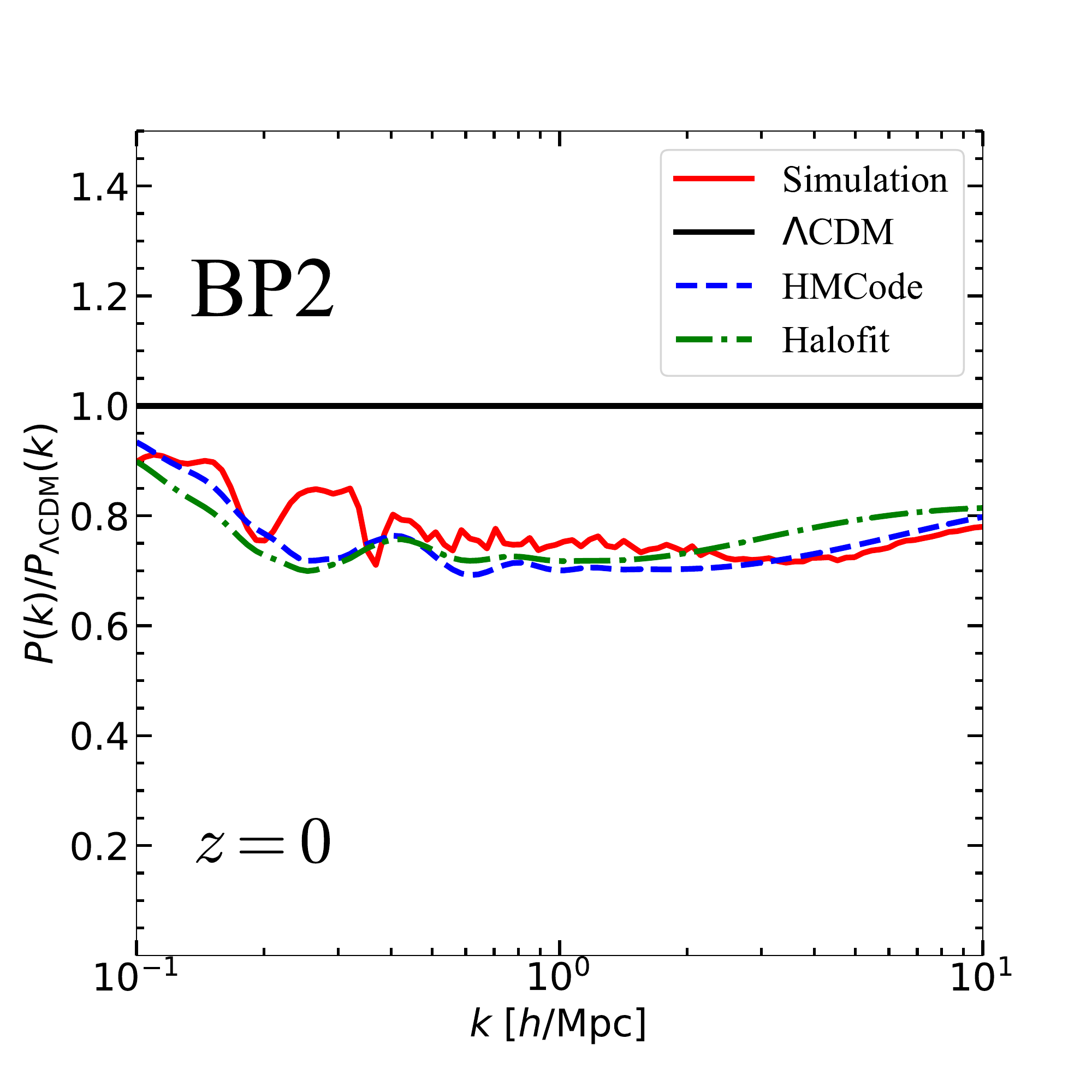}
    \includegraphics[scale = 0.37]{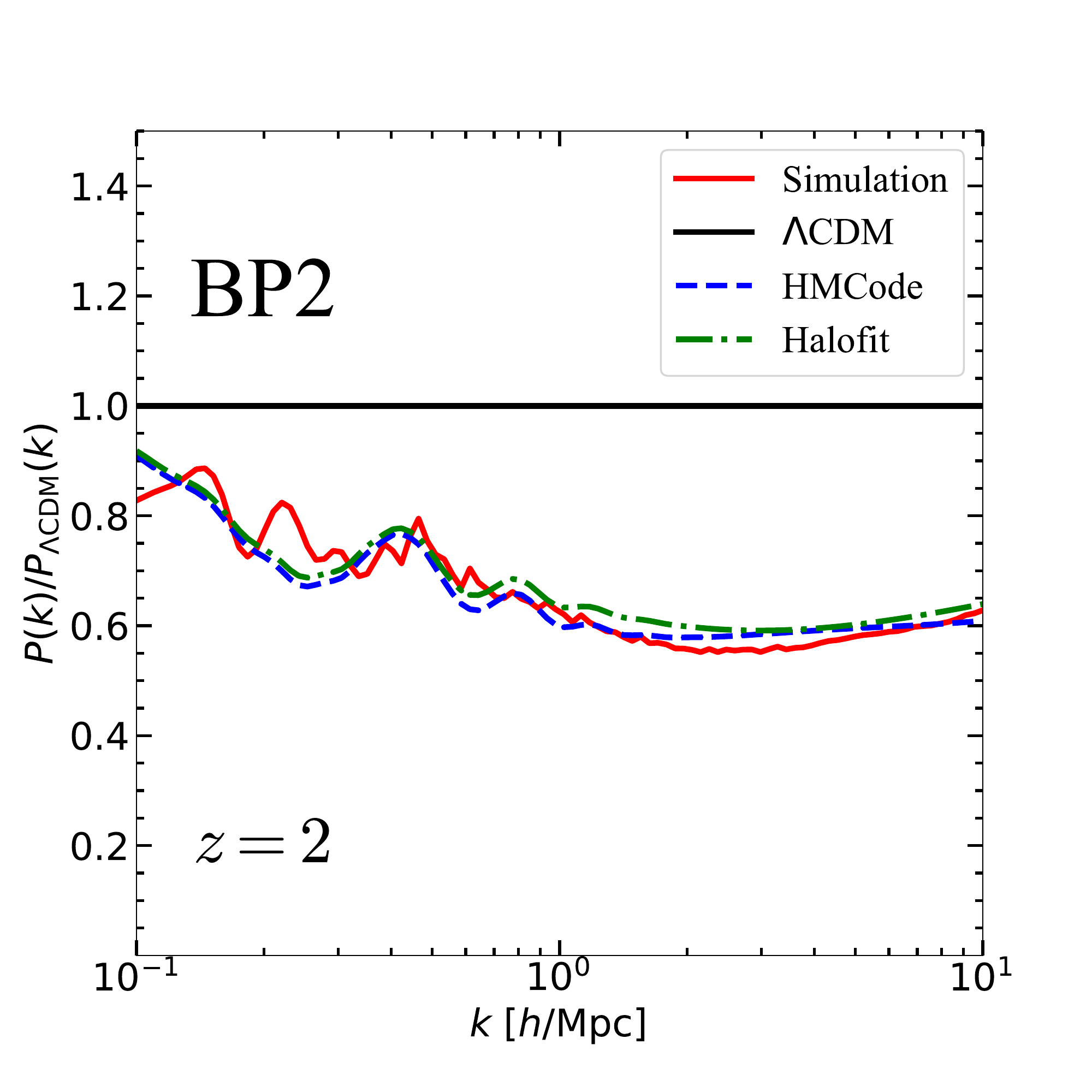}
    \caption{The ratio of matter power spectrum between MTH model and $\Lambda \rm{CDM}$. 
    The numerical results are computed by  
    the public nonlinear codes (\texttt{Halofit} and \texttt{HMCode}) and 
    $N$-body simulation at $z = 0$ (left panel) and $z = 2$ (right panel) for \textbf{BP1} (upper panel) and \textbf{BP2} (lower panel). 
    The red solid lines are the results obtained by the $N$-body simulation, while 
    the green dashed-dotted lines and the blue dashed lines correspond to the results given by  
    \texttt{Halofit} and \texttt{HMCode}, respectively. }
    \label{fig:simulation}
\end{figure}

We perform our $N$-body simulations with the code \texttt{P-Gadget3}, a TreePM code based on the publicly available code \texttt{Gadget2}~\cite{Springel:2005mi}. 
We start our simulation from the redshift $z = 127$ with the initial condition generated by 
the code \texttt{2LPTic}~\cite{Crocce:2006ve} whose input power spectrum can be obtained by using \texttt{CLASS}. 
We use the matter power spectra generated by two different MTH models in the study. The two  benchmark models, which we call {\bf BP1} and {\bf BP2}, have the MTH parameters \{$\hat{r} = 0.25$, $\hat{v}/v = 6.31$, $\Delta\hat N = 0.35$\} and \{$\hat{r} = 0.10$, $\hat{v}/v = 3.00$, $\Delta\hat N = 0.30$\}, and the other cosmological parameters relevant to the power spectrum calculation are
\begin{itemize}
    \item[\textbf{BP1}]: 
    \{$\Omega_{\rm m}= 0.2936$, $\Omega_\Lambda= 0.7064$, $h = 0.7084$, $n_{\rm s} = 0.9727$, $\sigma_8 = 0.7599$\},
    \item[\textbf{BP2}]:
    \{$\Omega_{\rm m}= 0.3254$, $\Omega_\Lambda= 0.6746$, $h = 0.6756$, $n_{\rm s} = 0.9727$, $\sigma_8 = 0.7648$\}.
\end{itemize}
In the gravity-only simulation we perform, the code only takes the total matter density $\Omega_{\rm m}=\Omega_{\rm cdm}+\Omega_b$. Since Gadget only reads the input power spectrum shape, we calculate the $\sigma_8$ of each model separately and feed them to Gadget to calibrate the power spectrum. We run the $N$-body simulation with $N_0=512^3$ particles in a periodic box of volume $(200~{\rm Mpc}/h)^3$. 
The mass of a simulation particle is $4.733\times10^9M_{\odot}/h$ in the simulations.  

In Fig.~\ref{fig:simulation}, we show the ratio of matter power spectra between the MTH and $\Lambda \rm{CDM}$ models obtained using the $N$-body simulation (red solid line), \texttt{Halofit} (green dash-dotted line), 
and \texttt{HMCode} (blue dashed line). We use the \texttt{Halofit} for the non-linear correction to the $P_{\Lambda{\rm CDM}}$ since the code has been tuned to mimic the $N$-body results of the $\Lambda$CDM model. As we can see from the red curves, although some features of the twin acoustic oscillations remain visible from the $N$-body simulation at $z=2$ (right panel), the features get smoothed out at $z=0$ (left panel), similar to the findings in Ref.~\cite{Schaeffer_2021,Munoz:2020mue}. The minor kinks at $k < 0.3~h/\rm{Mpc}$ come from  statistical fluctuations.

For $k\geq 1\,h$Mpc$^{-1}$, \texttt{HMCode} generate power spectra that agree well with the $N$-body simulation results. 
For $0.3\lesssim k\lesssim1\,h$Mpc$^{-1}$, the results deviate more due to the less severe damping of the oscillation pattern predicted by the \texttt{HMCode}. However, the difference to the $N$-body simulation is only up to $\approx 20\%$ for the BP1 with $k$-mode around $k=0.4\,h$Mpc$^{-1}$. The deviation is comparable to the fractional uncertainties $\approx 10\%$ in the DES~Y3 data. 

Given that the \texttt{HMCode} reproduces the general behavior of the $N$-body simulation results in these examples, and the oscillation patterns only remain in a small $k$ window and have amplitudes comparable to the experimental uncertainty, we believe the code can provide a reasonable estimate of the MTH spectrum for our MCMC study. Since the $N$-body simulation is much more time-consuming, using \texttt{HMCode} in the analysis is the best of what we can do now for the study. However, different from the \texttt{HMCode}, \texttt{Halofit} produces a much more different power spectrum in the \textbf{BP1} case at $z=0$, and it is hard to explain the origin of the deviation. We therefore do not use \texttt{Halofit} in the MCMC study.

\begin{figure}[t!]
    \centering
    \includegraphics[scale = 0.37]{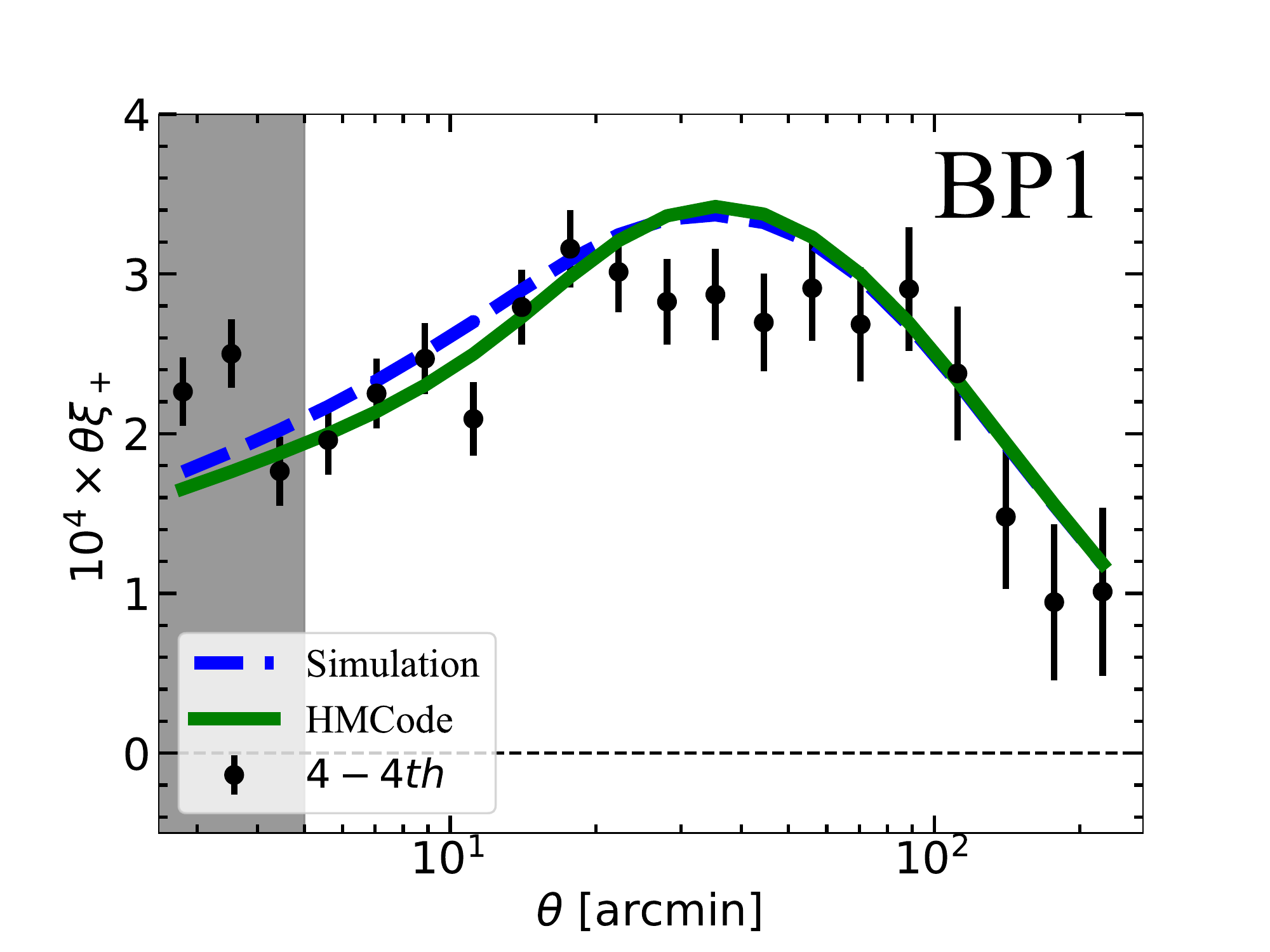}
    \includegraphics[scale = 0.37]{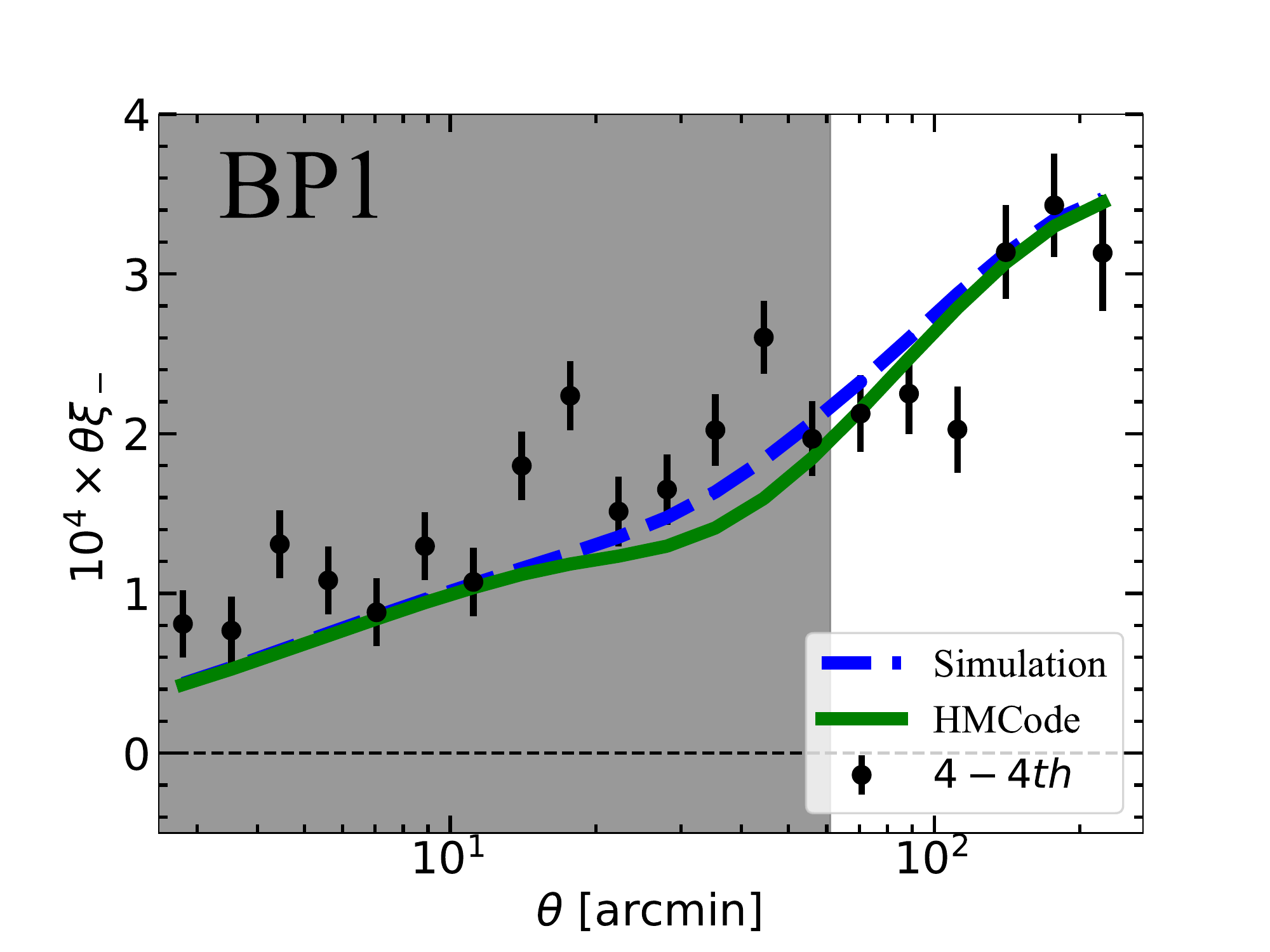}
    \includegraphics[scale = 0.37]{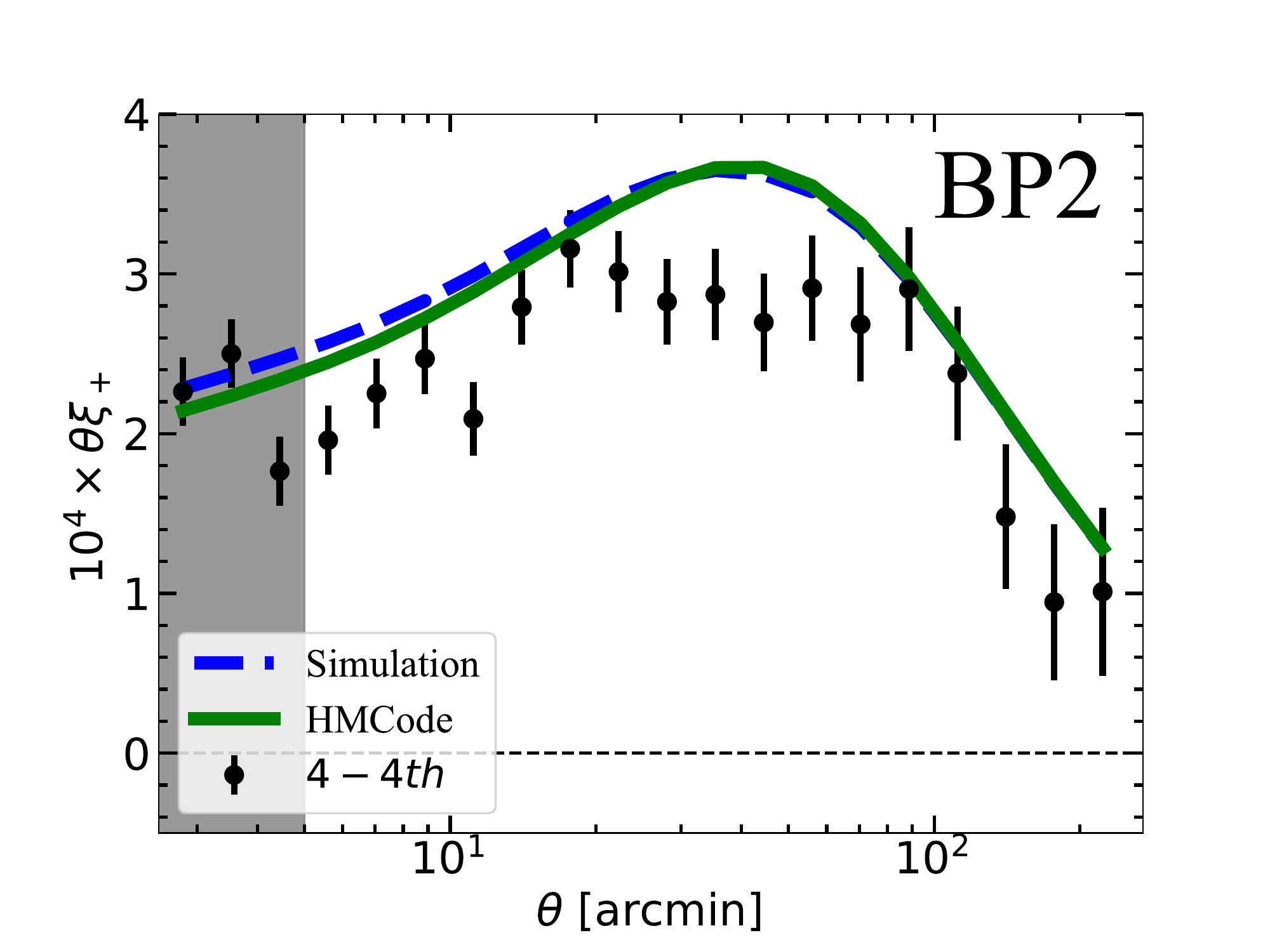}
    \includegraphics[scale = 0.37]{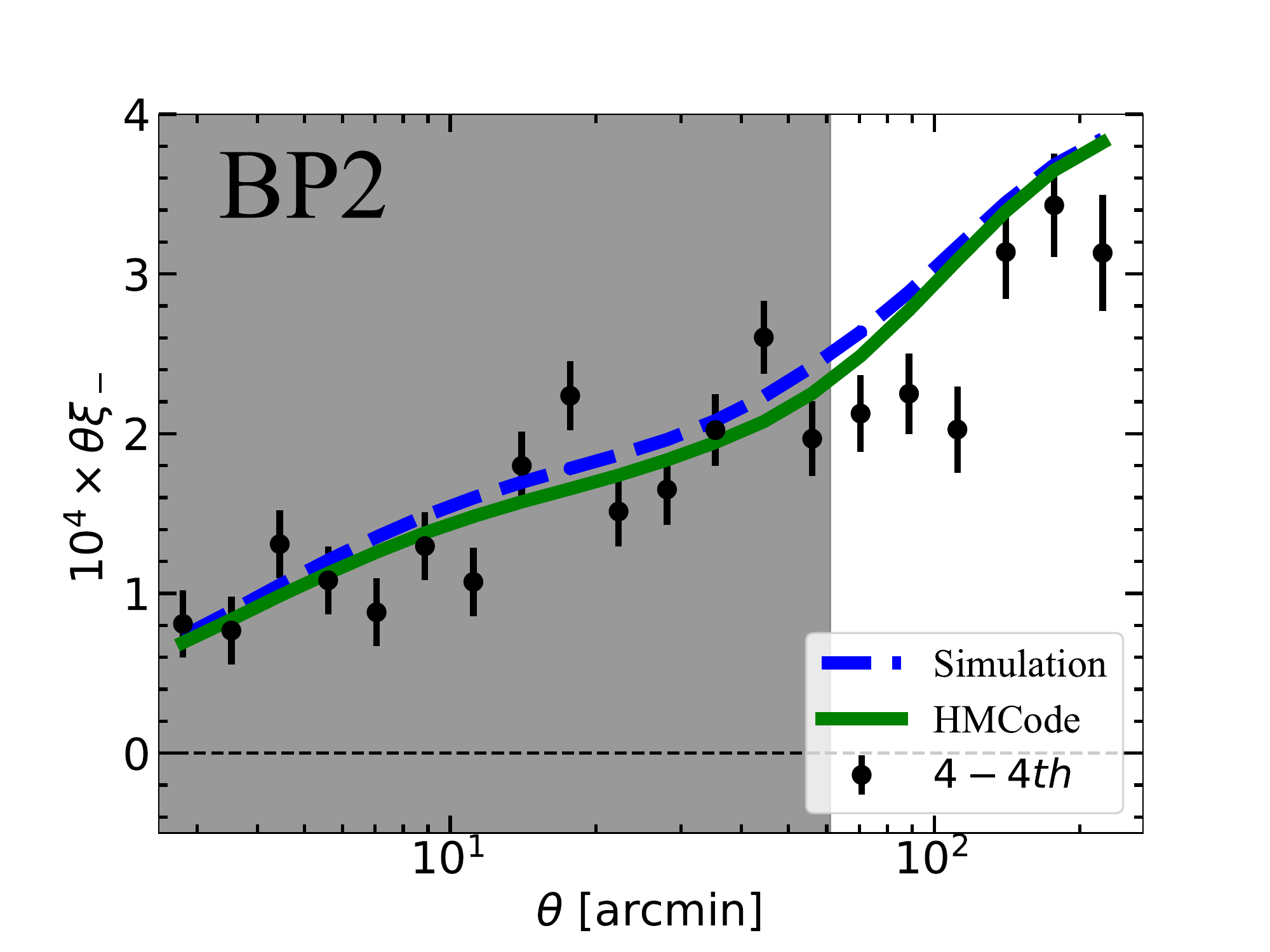}
    \caption{The cosmic shear of the 4-4${th}$ redshift bin obtained for the {\bf BP1} (upper) and {\bf BP2} (lower) models. 
    The data points correspond to the \texttt{DES~Y3} cosmic shear measurement. 
    The diagonal term of the covariance matrix gives the error bars of data points. The gray region is masked to avoid the small scale uncertainties as applied in Ref.~\cite{DES:2021bvc}. 
    The blue dashed lines correspond to the calculation based on the $N$-body simulation correction, 
    while the green solid lines correspond to the calculation based on the \texttt{HMCode}. 
    The difference between the $N$-body simulation and \texttt{HMCode} calculation is within the size of the error bars and mainly shows up within the masked (gray) region. 
    }
    \label{sim_err}
\end{figure}


To better quantify the effect of the deviation between the \texttt{HMCode} and $N$-body simulations to the weak lensing spectrum, 
we show the cosmic shear two-point correlation function for the 4-4${th}$ redshift bin in Fig.~\ref{sim_err}. 
Based on the same benchmark parameters but now with \{$\Omega_b h^2 = 0.02283$, $\Omega_{\rm cdm} h^2 = 0.1245$\} for \textbf{BP1} and 
\{$\Omega_b h^2 = 0.0227$, $\Omega_{\rm cdm} h^2 = 0.1258$\} for \textbf{BP2},
we plot the simulation result with blue dashed lines, 
while the predictions of \texttt{HMCode} are given by the green solid lines.  
We can see that the deviation is reasonably small compared to the size of the error bars.
In terms of the total $-\mathcal{L}$ of the fit to the DES~Y3 data, 
we find that the \texttt{HMCode} and $N$-body simulation results differ by $-2\ln(\mathcal{L}_{\rm MTH,HM}/\mathcal{L}_{{\rm MTH},N-{\rm body}})\approx3$ for {\bf BP1} and $\approx 30$ for {\bf BP2}. Given that the DES data has in total 227 degrees of freedom after the mask, the deviation is not significant. In turns of the MTH parameters, we can raise the green curve (\texttt{HMCode}) in Fig.~\ref{sim_err} to match the blue curve ($N$-body) by reducing $\hat r=0.25$ ({\bf BP1}) to $0.21$ and $\hat r=0.1$ ({\bf BP2}) to $0.07$ approximately. The  $\approx 30\%$ correction to the MTH parameter does not change the main results in this work. Therefore, although a precise analysis requires the $N$-body simulation, 
we believe that \texttt{HMCode} provides a valid nonlinear correction with tolerable error in the study.

\section{Summary and conclusion} 
\label{sec:sum}
In this study, we improve the cosmological constraints on the MTH model, which offers a solution to the Higgs little hierarchy problem, by incorporating DES 3-year cosmic shear data into our analysis. Our MCMC analysis, incorporating \texttt{Planck+BAO+DES Y3} likelihood, reveals that the DES Y3 cosmic shear data impose significantly tighter constraints on the model, necessitating a twin baryon composition fraction $\hat{r}\lesssim0.3$ ($2\sigma$) when $\Delta\hat N>0.05$ ($T_{\hat\gamma}>0.29\,T_{\gamma}$). The DES data disfavor a significant suppression in the matter power spectrum between $0.1\lesssim k\lesssim 1\,h$Mpc$^{-1}$, providing stronger bounds compared to the previous result in~\cite{Bansal:2021dfh}. 

When fitting the CMB+BAO data with the MTH model, the additional radiation in the twin sector and the presence of the twin BAO process can simultaneously increase the $H_0$ value and decrease the derived matter power spectrum, providing an opportunity to alleviate the $H_0$ and $S_8$ tensions. 
In contrast, most models that address the $H_0$ tension make the $S_8$ problem worse~\cite{Chacko:2016kgg,Lesgourgues:2015wza,Pandey:2019plg}. By incorporating the \texttt{SH0ES+DES Y3} data, we demonstrate that the MTH model can alleviate the SH0ES tension to a similar degree as the $\Lambda$CDM~$+\Delta N_{\rm eff}$ model, while simultaneously achieving a lower $S_8$ value that is more consistent with the DES~Y3 result. 

If future measurements of $S_8$ yield a similar mean value but with a reduced error bar of half the size of the DES~Y3 result, the resulting number would be comparable to the Planck~SZ~(2013) result, but with a significant $S_8$ tension to the CMB measurement. To consider the scenario with both significant $H_0$ and $S_8$ tensions, we include \texttt{SZ~(2013)+SH0ES +DES~Y3} in the analysis. Our result shows that the MTH model's best-fit parameters significantly improve the fits with only three additional parameters compared to the $\Lambda$CDM model, resulting in a decrease of likelihood $-2\ln(\mathcal{L}_{\rm MTH}/\mathcal{L}_{\Lambda{\rm CDM}})\approx-15$. The primary improvement comes from fitting the Planck~SZ (2013) data, with  $-2\ln(\mathcal{L}_{\rm MTH}/\mathcal{L}_{\Lambda{\rm CDM}})_{\rm SZ}=-15.3$, while the fit to the SH0ES data also shows a slight improvement with  $-2\ln(\mathcal{L}_{\rm MTH}/\mathcal{L}_{\Lambda{\rm CDM}})_{\rm SH0ES}=-3.3$. The MTH model is more responsive to the suppression of the local matter power spectrum than to the peak scale in CMB measurements, although it does help to increase $H_0$ while simultaneously decreasing $S_8$. Our fit to \texttt{Planck+BAO+SZ (2013)+SH0ES+DES~Y3} yields a $95\%$ credible region with non-zero values of $0.02 \lesssim \hat{r} \lesssim 0.4$ and $0.03 \lesssim \Delta\hat{N} \lesssim 0.5$.

Assuming that the MTH model is the correct description of the universe, we evaluate the expected precision of the future CSST experiment in determining the MTH parameters. With its larger effective area and greater number of observed galaxies, the CSST has the potential to reduce the uncertainties in the cosmic shear measurement by up to an order of magnitude. In Fig.~\ref{fig:r_dn_h0_s8_sz_csst}, we show that the CSST can significantly improve the precision of the MTH measurement by narrowing the fractional uncertainty of twin baryon energy density from an $\mathcal{O}(1)$ factor to approximately $10\%$.

Lastly, we address potential systematic uncertainties stemming from the differences between the \texttt{HMCode} and $N$-body simulations. In our MCMC analysis incorporating DES Y3, we must account for the non-linear correction to the matter power spectrum. To assess the validity of using \texttt{HMCode}, we compared a few examples of the MTH power spectra obtained from \texttt{HMCode} and a gravity-only $N$-body simulation (\texttt{P-Gadget3}). Our results indicate that \texttt{HMCode} provides an adequate description of the non-linear correction for the MTH model. The corrections to the lensing spectra resulting from our study are found to be smaller than the observational uncertainty. Additionally, based on the benchmark examples we examine, these corrections only correspond to a correction of $\mathcal{O}(10)\%$ to the twin baryon abundance. Thus, this study confirms the suitability of using \texttt{HMCode} for a comprehensive analysis.

\section*{Acknowledgments}
We thank Saurabh Bansal for providing the MTH module of the \texttt{CLASS} code and for his valuable assistance with the MCMC analysis.
We are grateful to Xiaoyuan Huang for supporting the computation resource. 
We are also grateful to Qiang Yuan, Jared Barron and Matthew Low for reading the manuscript and for providing useful comments. 
This work is supported by the National Key Research and Development Program of China (No. 2022YFF0503304), CSST science funding, 
the CAS Project for Young Scientists in Basic Research (YSBR-092), 
the National Natural Science Foundation of China (No. 11921003, No. 12003069, No. U1738210), and 
the Entrepreneurship and Innovation Program of Jiangsu Province. HZC acknowledges the funds of
the cosmology simulation database (CSD) in the National Basic Science Data Center (NBSDC). YT is supported by
the U.S. National Science Foundation (NSF) grant PHY-2112540.

\appendix
\section{Intrinsic alignment on weak lensing analysis}
\label{sec:IA}
One of the major astrophysical systematic uncertainty in the cosmic shear analysis comes from the intrinsic alignment (IA) of galaxy shapes. The shapes of galaxies exhibit correlations with their local environments that effect the galaxies' formation and evolutionary histories. 
In this work, we adopt a commonly used non-linear alignment model (NLA) 
to deal with IA~\cite{Hamana:2019etx,Hirata:2004gc,Bridle:2007ft,Joachimi:2010xb}. 
In principle, the 2D power spectrum can own three contributions, 
\begin{equation}
    C_\ell^{total} = C_{GG}^{ij}(\ell)+C_{GI}^{ij}(\ell)+C_{II}^{ij}(\ell).
\end{equation}
The later two terms receive  contributions from environmental interactions: (a) $C^{ij}_{II}$ denotes the correlation between intrinsic shapes of galaxies that are physically close to each other, and (b) $C^{ij}_{GI}$ denotes the cross shear-intrinsic correlations between galaxies on the neighbouring lines of sight. The GG term denotes the correlations of gravitational shear given by Eq.~\eqref{eq:cl_gg}. Th $II$ and $GI$ terms can be expressed as:
\begin{eqnarray}
C_{II}^{i j}(\ell)&=& \int_0^{\chi_H} d \chi F^2(\chi) \frac{n^i(\chi) n^j(\chi)}{\chi^2} P_\delta\left(\frac{\ell+1/2}{\chi}, z(\chi)\right)\,
\\
C_{GI}^{i j}(\ell)&=& \int_0^{\chi_H} d \chi F(\chi) \frac{W^i(\chi) n^j(\chi)+n^i(\chi) W^j(\chi)}{\chi^2} P_\delta\left(\frac{\ell+1/2}{\chi}, z(\chi)\right)\,,
\end{eqnarray}
where $F(z)$ denotes the correlation strength between the tidal field and the galaxy shapes
\begin{equation}
    F(z) = -A_{IA}C_1\frac{\Omega_m}{D(z)}\,\rho_{c}\left(\frac{1+z}{1+z_0}\right)^\eta\,.
\end{equation}
$A_{IA} \ \& \ \eta$ are the amplitude and index of redshift dependence respectively, both are free parameters. $\rho_c$ is the critical density at z=0.  $\Omega_m$ is the total matter abundance in the universe at z=0. $D(z)$ is the linear growth factor normalized to unity at z=0. $C_1 = 5\times 10^{-14} h^{-2} M_\odot^{-1} \rm{Mpc}^3$ is a normalized fixed constant and set the pivot redshift $z_0 = 0.62$.

\section{Supplementary Figures}
\label{sec:appendix_fig}

\begin{figure}[h]
    \centering
    \includegraphics[scale = 0.4]{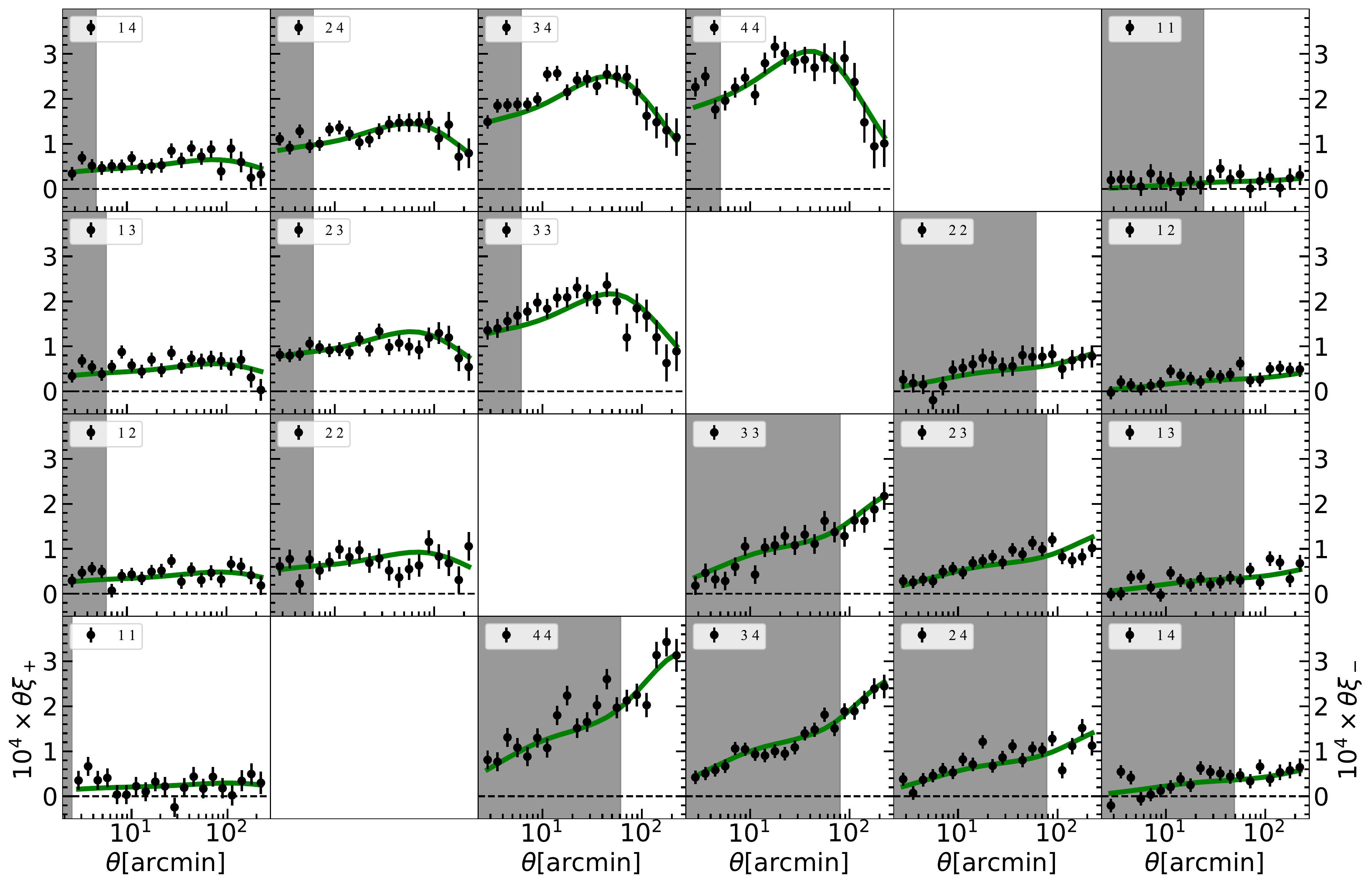}
    \caption{All auto correlations and cross correlations between different redshift bins. 
    The green lines are the MTH prediction and its  parameters are given 
    in the column \texttt{Planck+BAO+DES Y3+SZ~(2013)+SH0ES} 
    of Table~\ref{tab:bestfit}.
    }
    \label{xi_pm_tot}
\end{figure}

In Fig.~\ref{xi_pm_tot}, we show the full set of auto and cross two-point correlation functions of cosmic shear in each redshift bin. The top left corner is $\xi_+(\theta)$ while the bottom right corner is $\xi_-(\theta)$. The black data points in the plots are data from the DES~Y3 \texttt{METACALIBRATION} shape catalog. The green solid lines denote the best-fit MTH model prediction with dataset \texttt{Planck+BAO+DES Y3+SH0ES+SZ~(2013)} in Table~\ref{tab:bestfit}. The error bars are the diagonal term of the covariance matrix described in Sec.~\ref{covmat}. 

\begin{figure}[h]
    \centering
    \includegraphics[scale = 0.37]{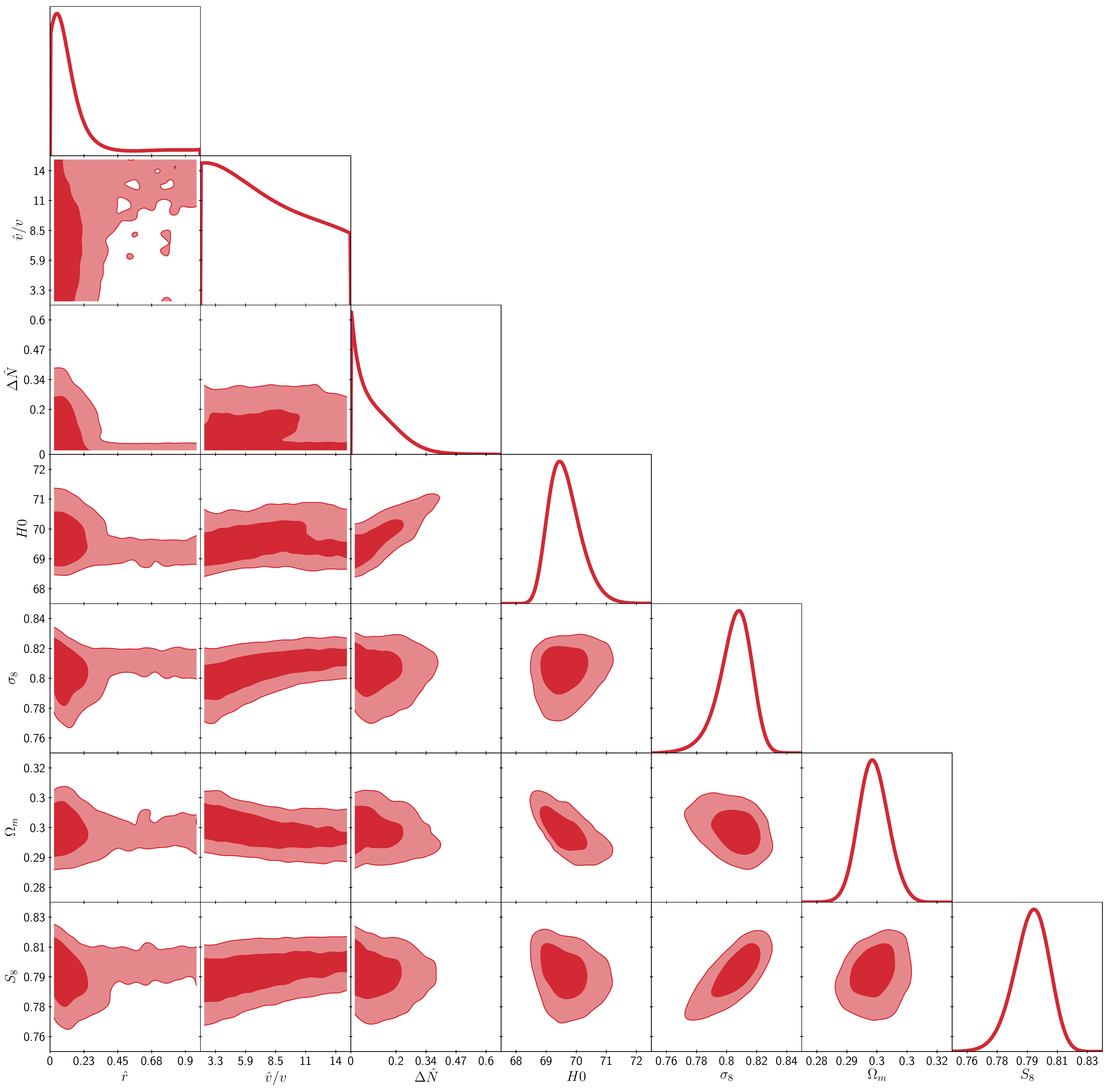}
    \caption{The triangle posterior distribution of MCMC results for the MTH model with dataset \texttt{Planck+BAO+DES Y3}.} 
    \label{reweight_corner}
\end{figure}

\begin{figure}[h]
    \centering
    \includegraphics[scale = 0.37]{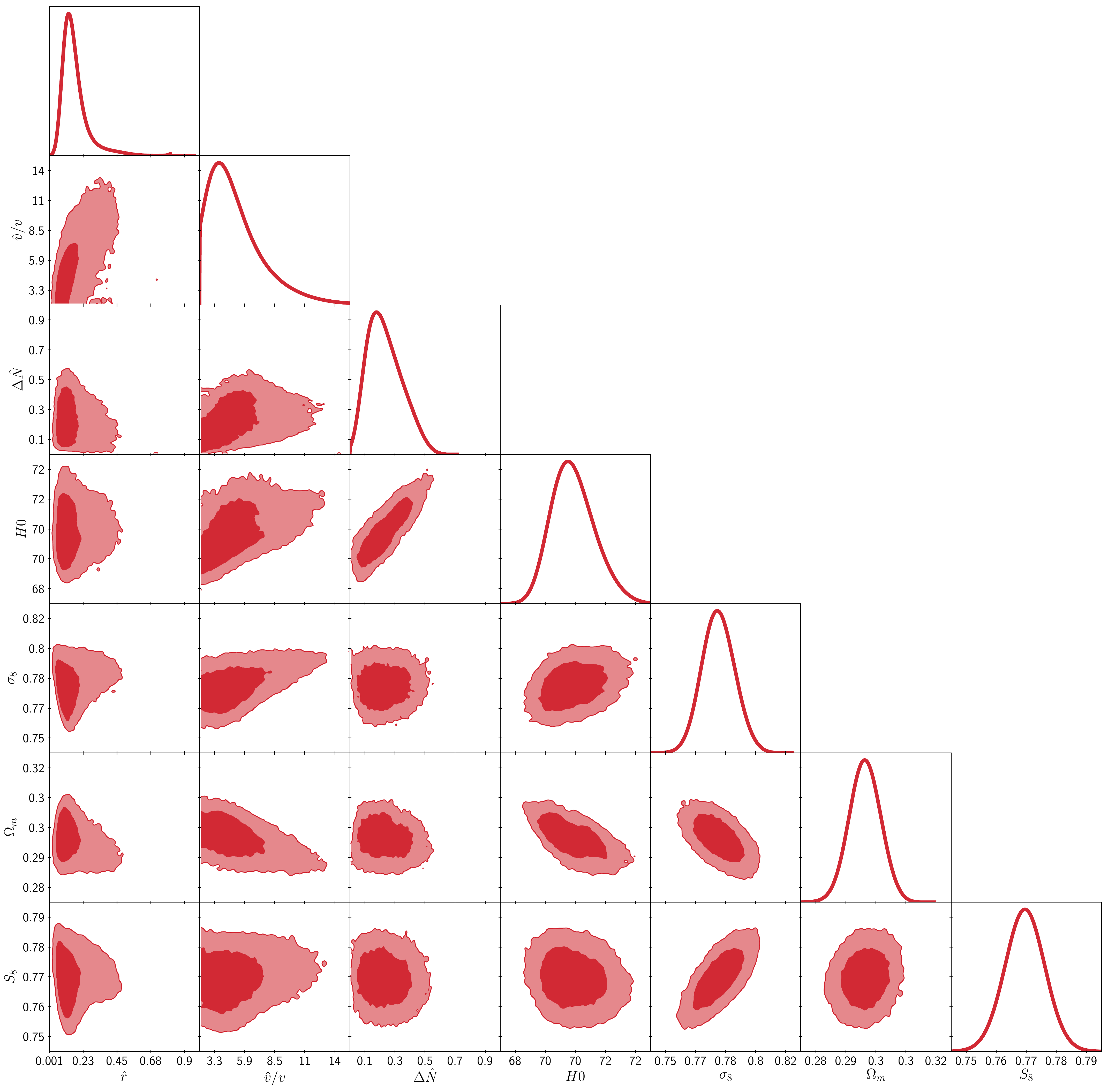}
    \caption{The triangle posterior distribution of MCMC results for the MTH model with dataset \texttt{Planck+BAO+DES Y3+SH0ES+SZ~(2013)} experiments.}
    \label{MTH_sz}
\end{figure}

In Fig.~\ref{reweight_corner} and Fig.~\ref{MTH_sz}, we show 
the triangle contour figures of the MTH model with datasets \texttt{Planck+BAO+DES Y3} and \texttt{Planck+BAO+DES Y3+SZ~(2013)+SH0ES} respectively.

\bibliographystyle{JHEP}
\bibliography{references}

\end{document}